\documentclass[12pt,preprint]{aastex}

\slugcomment{Draft}

\shorttitle{Chemical abundances of the Milky Way thick disk and stellar halo}
\shortauthors{Ishigaki, Aoki, and Chiba}

\begin{document}


\title{Chemical abundances of the Milky Way thick disk and stellar halo II.: sodium, iron-peak and neutron-capture elements}


\author{M. N. Ishigaki}
\affil{Kavli Institute for the Physics and Mathematics of the Universe (WPI), the University of Tokyo}
\affil{5-1-5 Kashiwanoha, Kashiwa, Chiba, 277-8583, Japan}
\email{miho.ishigaki@ipmu.jp}

%\affil{National Astronomical Observatory of Japan}
%\affil{2-21-1 Osawa, Mitaka, Tokyo 181-8588, Japan}
%\email{ishigaki.miho@nao.ac.jp}

\author{W. Aoki}
\affil{National Astronomical Observatory of Japan}
\affil{2-21-1 Osawa, Mitaka, Tokyo 181-8588, Japan}
\email{aoki.wako@nao.ac.jp}

\and

\author{M. Chiba}
\affil{Astronomical Institute, Tohoku University}
\affil{Aoba-ku, Sendai 980-8578, Japan}
\email{chiba@astr.tohoku.ac.jp}


%\altaffiltext{1}{Visiting Astronomer, Cerro Tololo Inter-American Observatory.
%CTIO is operated by AURA, Inc.\ under contract to the National Science
%\altaffiltext{2}{Society of Fellows, Harvard University.}
%\altaffiltext{3}{present address: Center for Astrophysics,
%    60 Garden Street, Cambridge, MA 02138}
%\altaffiltext{4}{Visiting Programmer, Space Telescope Science Institute}
%\altaffiltext{5}{Patron, Alonso's Bar and Grill}


\begin{abstract}

We present chemical abundance analyses of sodium, iron-peak 
and neutron-capture elements for 97 kinematically 
selected thick disk, inner halo and 
outer halo stars with metallicities $-3.3<$[Fe/H]$<-0.5$. 
The main aim of this study is to examine chemical similarities and 
differences among metal-poor stars belonging to
 these old Galactic components as a clue to determine their early 
chemodynamical evolution. In our previous paper,  
we obtained abundances of 
$\alpha$ elements by performing a one-dimensional LTE abundance analysis
 based on the high-resolution 
($R\sim 50000$) spectra obtained with the Subaru/HDS. 
In this paper, a similar analysis is performed 
to determine abundances of an additional 17 elements. 
We show that, in metallicities below [Fe/H]$\sim-2$, 
the abundance ratios of many elements in the 
thick disk, inner halo, and outer halo subsamples 
are largely similar. 
In contrast, in higher metallicities ([Fe/H]$\gtrsim -1.5$), 
differences in some of the abundance ratios among the three 
subsamples are identified. Specifically, 
 the [Na/Fe], [Ni/Fe], [Cu/Fe], and [Zn/Fe] ratios 
in the inner and outer halo subsamples are 
found to be lower than those 
in the thick disk subsample. 
A modest abundance difference between the two halo
subsamples in this metallicity range 
is also seen for the [Na/Fe] and [Zn/Fe] ratios. 
In contrast to what was observed for [Mg/Fe] in our previous paper, 
[Eu/Fe] ratios are more enhanced in the two halo subsamples 
rather than in the thick disk subsample. 

The observed distinct chemical abundances of some elements 
between the thick disk and inner/outer halo subsamples with [Fe/H]$>-1.5$ 
support the hypothesis that these components formed through different mechanisms. 
In particular, our results favor the scenario that the 
inner and outer halo components formed through an assembly of 
multiple progenitor systems that experienced various degrees of 
chemical enrichments, while the thick disk formed through
rapid star formation with an efficient mixing of chemical elements. 
The lower [Na/Fe] and [Zn/Fe] observed in stars with the outer halo 
kinematics may further suggest that progenitors with longer 
star formation timescales contributed to the build-up of the relatively 
metal-rich part of stellar halos.

\end{abstract}





\section{Introduction}

Due to the advent of dedicated photometric and spectroscopic 
surveys such as the Sloan Digital Sky Survey (SDSS), 
the old metal-poor components of our Milky Way (MW) Galaxy, namely, 
the thick disk and the stellar halo, are found to 
be far more complex than previously thought exhibiting 
various degrees of substructures. The origins of 
the complex nature of these old structural components have been 
studied through observations and theoretical models as 
one of the central issues to unveil how the MW formed 
and evolved along with the evolution of galaxies in the universe.  
Although these efforts were successful in greatly advancing our view of 
these old MW components, their origins remain unclear.

Although the MW thick disk has been known for many years, 
its dynamical and chemical structures are still controversial to 
constrain its formation mechanism.
The MW thick disk was originally 
discovered as an extra component required 
to fit the stellar number density distribution over 
the thin disk component at distances from the disk plane 
greater than $\sim 1$ kpc \citep{yoshii82,gilmore83}. 
The thick disk is widely believed to have been formed 
at an early epoch of Galaxy formation because constituent 
stars have older ages and lower metallicities compared to 
the thin disk stars \citep[e.g.,][]{bensby03}. 
In the solar neighborhood, thick disk stars lag in rotational velocities
 behind thin disk stars by $\sim 20-50$ km s$^{-1}$ \citep{chiba00,carollo10}. 
Chemical abundances of the thick disk stars are known to 
be characterized by higher [$\alpha$/Fe] ratios than 
the thin disk stars \citep{fuhrmann98,bensby03,
bensby05,reddy03,reddy06,prochaska00,ruchti11}, more similar to 
the stars in the Galactic bulge \citep{melendez08,bensby10}.
Systematic analyses of low-resolution spectra
for a large number of stars from the Sloan Extension 
for Galactic Understanding and Exploration (SEGUE) 
have significantly advanced our view of this component 
\citep[e.g.,][]{carollo10,lee11,cheng12}.
In particular, based on the SDSS/SEGUE data, 
\citet{carollo10} suggest that the previously 
known metal-poor tail ($-2.7<$[Fe/H]$<-1.0$) of the metallicity 
distribution function 
of the thick disk, which is called the ``metal-weak thick disk'' (MWTD; 
\citet[e.g.,][]{beers02}), could indeed be associated with an 
independent stellar component. 
On the other hand, \citet{bovy12} questioned the distinct nature 
of the thick disk apart from the thin disk based on the 
analysis of chemically defined subpopulations by carefully 
taking into account possible selection biases in the SDSS 
spectroscopic data.  

The MW stellar halo is known as another oldest remnant 
of the early chemodynamical evolution of our Galaxy. 
Recent large surveys such as SDSS reveal 
that the MW stellar halo is highly structured and cannot be 
approximated by a smooth 
spherical distribution of metal-poor stars \citep[e.g.,][]{ivezic12}. 
It became clear that the stellar halo contains a number of 
substructures in spatial distribution 
\citep[e.g.,][]{ibata94,newberg02,majewski03,juric08} or kinematics \citep[e.g.,][]{helmi99,chiba00,kepley07,schlaufman09} of constituent stars. 
The most prominent example is a currently merging dwarf galaxy, Sagittarius 
and its tidal debris \citep[e.g.][and reference therein]{helmi08,ivezic12}. 
Furthermore, \citet{carollo07,carollo10} recently confirmed 
that the MW stellar halo is divisible into two globally overlapping 
stellar components, 
namely, the inner and outer stellar halos, based on the analysis 
of kinematics and metallicity for a large sample of 
calibration stars obtained by SDSS/SEGUE.
These authors showed that the inner halo dominates 
at Galactocentric distances $r$ smaller than $\sim 10-15$ kpc. It has 
a modestly flattened distribution of stars with a mean 
metallicity of [Fe/H]$\sim -1.6$. 
The kinematics of the stars likely belonging to the inner halo 
are characterized by a zero to slightly prograde mean 
rotational velocity with large velocity dispersions. 
In contrast, the outer halo was found to 
dominate at $r>10-15$ kpc and exhibit a more spherical 
stellar density distribution. The mean metallicity was 
estimated to be [Fe/H]$\sim -2.2$, which is lower than 
that of the inner halo. 
The kinematics of the outer halo component 
are characterised by a larger number of 
extreme motions (e.g. large prograde 
or retrograde orbit) that 
are the outlier of the typical inner halo stars 
\citep{carollo07,carollo10}. Although the definite rotational 
properties of the two components remain under discussion
\citep[e.g.,][]{schonrich11,beers12}, 
the proposed difference in the structural properties between the 
inner and outer parts of the MW halo suggests that 
these two components formed through different mechanisms 
\citep{carollo10}.

Chemical abundances of individual metal-poor stars that constitute 
the thick disk and stellar halos provide unique opportunity to 
test theoretical models for the chemodynamical evolution of these 
components, 
particularly for the possible progenitors of these 
systems \citep{freeman02}.  
It has long been known that chemical abundances of metal-poor stars 
in the Solar neighborhood are characterized by an enhancement 
of $\alpha$ elements (Mg, Si, Ca, and Ti) to iron 
abundance ratios ([$\alpha$/Fe];
\citet[e.g.,][]{luck81,mcwilliam95,cayrel04,lai08}).  
This result is usually interpreted as evidence that
the MW field metal-poor stars are 
formed out of gas enriched mainly through Type II Supernovae (SNe) of 
massive stars, 
which would yield high [$\alpha$/Fe] \citep[e.g.,][]{matteucci86}. 
On the other hand, observations of spasial
distribution, kinematics, and metallicity for a number of 
MW stars suggest that 
the metal-poor stars in the Galaxy are a mixture of 
apparently different stellar populations \citep[e.g.,][]{bell08}. 
If each of these populations has their own chemical enrichment 
histories, we would expect to distinguish these populations 
through kinematics and detailed chemical abundances of individual 
metal-poor stars.

In this context, any correlation between orbital kinematics 
and chemical compositions is of particular interest 
since a star with extreme kinematics 
is more likely to have been accreted from an external system, 
which may have quite different chemical enrichment histories 
in their birth place. Detailed chemical abundances 
of such stars provide us with clues for understanding properties such as
 star formation rates, efficiency of galactic wind, 
initial mass function (IMF), etc., of 
possible building blocks of our Galaxy \citep{lanfranchi03}.  
Taking advantage of a high-resolution spectrograph on 
8-10 m class telescopes, comprehensive studies on the 
correlation between detailed chemical abundances 
and kinematics of nearby halo stars have been carried out 
 \citep[][hereafter Paper I]{stephens02,fulbright02,roederer08,zhang09,ishigaki10,nissen10,
nissen11,ishigaki12}. 
\citet{stephens02} reported that stars having a  
large apocentric distance ($R_{\rm apo}$) tend to have 
lower [$\alpha$/Fe] while no correlations of 
abundances on other orbital parameters were identified. Since sample 
stars with extreme outer halo kinematics were still 
limited, the presence or absence of the correlation 
remained uncertain. 
Recently, \citet{nissen10} suggest that 
their sample of nearby dwarf stars can be divisible 
into two distinct groups in terms of abundances of 
several $\alpha$ elements, namely, the low-$\alpha$ 
and high-$\alpha$ stars. The two chemically distinct
groups of stars tend to show different characteristic kinematics. 
\citet{nissen11} further reported that these groups show 
different abundances in other elements such as  
Na and Zn.  These studies imply that the stars more metal-poor 
than [Fe/H]$<-0.5$ cannot be formed within a single well mixed 
gas but more likely formed in different pre-Galactic clumps 
that have their own chemical enrichment history.

It remained uncertain, however, how the presence of 
the chemically distinct groups of stars in the solar neighborhood 
is fit into the formation scenario for the thick disk
and inner/outer stellar halos. A sample of stars with a wide 
range of kinematics and metallicities, including those characteristics 
of the thick disk, inner halo, and outer halo components, 
 would provide useful insights into this issue.  
In this paper, we investigate similarities and differences in 
 detailed chemical abundance patterns 
among the kinematically selected thick disk, inner halo, and outer halo stars. 
This allows us to investigate whether or not the stars in these components with 
their overlapping metallicity range formed under the influence of a similar 
chemical enrichment history or not. 
 For this purpose a sample of 97
metal-poor ([Fe/H]$<-0.5$) stars spanning a wide range of orbital 
parameters and [Fe/H] are studied. In Paper I, 
we present the abundance analysis of  $\alpha$-elements (Mg, Si, Ca and 
Ti) for the sample stars. 
We showed that in a metallicity range of $-1.5<$[Fe/H]$<-0.5$, 
kinematically defined thick disk stars have higher [Mg/Fe] and [Si/Fe] 
with small scatter, while the inner and outer halo stars 
show lower average abundance ratios for these elements with larger scatter. 
In the present study, we further 
present the results for sodium, iron(Fe)-peak, and neutron-capture 
elements for the same sample of stars and investigate 
possible scenarios that consistently explain the observed abundance 
patterns in each of the old Galactic components.

This paper is organized as follows. In Section \ref{sec:sample}, we
 present a summary of our sample stars
and their membership to the thick disk, inner halo, and outer 
halo subsamples, that have been taken from Paper I. 
In Section \ref{sec:analysis}, a brief review of the observation, which 
was fully described in Paper I, is presented.  Then, 
determination of stellar atmospheric parameters and 
abundance analyses of individual elements are described. 
In Section \ref{sec:results}, 
we present the abundance results on the distribution in 
the [X/Fe]-[Fe/H] plane for each of the kinematically selected 
subsamples and examine 
the correlation between [X/Fe] and orbital parameters.   
Finally, Section \ref{sec:discussion} discusses the interpretation 
of the abundance differences and similarities among the three 
subsamples and their implications for the formation of these 
components.

\section{The sample}
\label{sec:sample}
The sample of 97 dwarf and giant stars with [Fe/H]$\le -0.5$ 
was selected from 
the catalogs of \citet{carney94}, \citet{ryan91}, and 
\citet{beers00} based on their kinematics. 
The proper motions and distance estimates 
were partly updated from those in the original catalog as described in 
Paper I. 
The radial velocities for the sample stars were also updated 
to those measured from the high-resolution spectra obtained in 
our observation. The orbital parameters $R_{\rm apo}$ (apocentric distance), 
$Z_{\rm max}$ (maximum distance from the Galactic plane), 
and $e$ (orbital eccentricity) were calculated by  
adopting the St\"{o}chel-type Galactic potential in the same manner as  
described in \citet{chiba00}. More details on the sample 
selection and kinematics of the sample stars 
are described in Paper I.

\subsection{Kinematics and membership assignment}
Based on the kinematics, we assigned the 
membership for the thick disk, inner halo or outer 
halo components to each of the sample stars as described in Paper I. 
In short, we calculated the  
probabilities that each of the sample stars belongs to  
the thick disk ($P_{\rm TD}$), inner halo ($P_{\rm IH}$) 
or outer halo components ($P_{\rm OH}$), based on their 
space velocities in the Galactic cylindrical 
coordinate ($V_{\rm R}$, $V_{\phi}$, and $V_{\rm Z}$). 
In this calculation, the mean velocities and dispersions 
for the thick disk, inner halo and outer halo 
components as well as a fractional contribution of each component 
at different $Z_{\rm max}$ were adopted from the values obtained by 
\citet{carollo10} based on their analyses of SDSS DR7. 
Then, the thick disk, inner halo, and outer halo stars were defined 
as the stars with $P_{\rm TD}>0.9$, $P_{\rm IH}>0.9$, and $P_{\rm OH}>0.9$, 
respectively. 
Other stars with $P_{\rm TD}, P_{\rm IH}, P_{\rm OH}\le 0.9$ 
were classified as either the thick disk/inner halo 
or the inner halo/outer halo intermediate populations. 
In the remainder of the paper, we conventionally refer to the stars classified 
as the thick disk, inner halo, and outer halo categories as the 
``thick disk, inner halo, and outer halo stars/ subsamples", respectively.
In the above definitions, 12, 34, and 37 stars are assigned to 
the thick disk, inner halo and outer halo components, respectively.
The adopted criteria for the membership assignment 
are purely based on kinematics while 
metallicities for each component are not taken 
into account. We briefly describe its consequence 
in the next subsection. 

As shown in Figure 1 of Paper I., 
the thick disk stars in our sample have a 
mean rotational velocity $V_{\phi}\sim 180$ km s$^{-1}$, whose 
orbit is confined to $\sim 1$ kpc above and below the Galactic 
plane ($Z_{\rm max}<1$ kpc). The inner halo stars show
no rotation $V_{\phi}\sim 0$ on average and exhibit larger 
velocity dispersion. Finally, the outer halo stars 
show a much larger dispersion in 
$V_{\phi}$ and some have extreme 
prograde or retrograde rotation. At the same time, 
almost all stars with orbits that reach the distance of 
$>10$ kpc from the Galactic 
plane were classified as the outer halo stars by definition.

\subsection{Metallicity of the sample}
\label{sec:metallicity}

Our classification of the sample stars into the 
thick disk, inner halo, and outer halo subsamples described 
above is purely based on kinematics. As a result,  
 metallicities for each of the three subsamples may be different 
from those of the thick disk, inner halo, 
and outer halo components obtained in previous works \citep[e.g.,][]{carollo10}. 
In this subsection, we compare 
metallicities of our thick disk, 
inner halo and outer halo subsamples with the 
metallicity distribution functions reported in the literature for 
each Galactic component.

The thick disk stars in our sample 
span a lower metallicity range than that 
reported for the canonical thick disk component and likely include 
stars with chemical and kinematical properties similar to the 
MWTD component.
The canonical thick disk component was reported to 
dominate in the metallicity range $-1.0\lesssim$[Fe/H]$\lesssim-0.4$ \citep[e.g.][]{wyse95}, 
while our thick disk subsample extends to the metallicity 
as low as $-2.7$. 
At metallicities below [Fe/H]$<-1.0$, \citet{carollo10} 
suggested that the independent MWTD component 
is required to account for the observed 
[Fe/H] and $V_{\phi}$ distribution for their sample stars close to 
the Galactic plane. 
They also reported that the metallicity of the 
MWTD component is in a range $-1.7<$[Fe/H]$<-0.8$ and 
$V_{\phi}\sim 100-150$ km s$^{-1}$ with a dispersion of 
$\sim 35-45$ km s$^{-1}$. 
Four of the thick disk stars in our sample with [Fe/H]$<-0.8$ have 
rotational velocities similar to the MWTD ($140<V_{\phi}<190$ km s$^{-1}$), 
where the two of the most metal-poor ones
have the lowest $V_{\phi}$ values. It is unclear 
whether these two stars represent the lowest metallicity tail 
of the MWTD or interlopers from the halo component.
 
Chemical abundances other than iron
for the MWTD have been investigated by several studies, in which 
a distinct chemical signature for this component was not clearly 
identified.
\citet{reddy08} studied elemental abundances for $\alpha$, iron-peak 
and neutron capture elements in the 14 candidate MWTD stars and found that 
their abundances are indistinguishable from halo stars with 
similar metallicity. \citet{ruchti11} studied abundances of iron 
and $\alpha$-elements for a large sample of metal-poor stars 
based on the medium-resolution spectroscopic data from 
the Radial Velocity Experiment. They reported that the metal-poor 
thick disk stars are enhanced in [$\alpha$/Fe] ratios similar 
to the halo stars. 

In Paper I, we investigate [$\alpha$/Fe] for the four stars 
with similar properties as the elusive MWTD as mentioned above. 
As a result, these stars were shown to have higher 
[Mg/Fe] or [Si/Fe] ratios than the inner/outer halo stars. 
Although the number of stars in this sample is very small to extract a definite 
conclusion about the properties of the proposed MWTD, 
we later compare their chemical abundances other than $\alpha$ elements 
with those of the typical thick disk, inner halo, and 
outer halo stars in our sample. 

Our classification of inner and outer-halo stars may not be 
representative of the stellar halo components observed in other 
surveys \citep{carollo10, dejong10, an13}. 
Stars with outer halo like kinematics are classified as outer halo 
members regardless of their [Fe/H]. Following the same criteria, 
stars with inner halo kinematics are assigned to the inner halo 
component independently from their metallicity. 
As a result, our sample of inner halo and outer halo stars both
span the wide range in metallicity $-3.5<$[Fe/H]$<-0.4$.
This is in contrast to the inner/outer halo division 
reported by \citet{carollo10}, where the inner halo stars have a  
peak metallicity of $-1.6$ while the outer halo stars 
have $\sim -2.2$ dex. 
Note that some 
of the stars assigned to the inner-halo or the outer-halo have 
metallicity in agreement with these components as derived in 
previous works. 

The reason for this metallicity difference is not clear. 
One possible explanation is that our sample selection in the 
solar-neighborhood ($<1-2$ kpc) may be biased toward/against particular 
metallicity among the inner halo or outer halo components. 
In the following, we focus on comparing 
abundance ratios between kinematically defined subsamples 
at a given metallicity.

\section{Observation and data analysis}
\label{sec:analysis}

\subsection{Observation and data reduction}
The observations for all of the sample stars were 
made with the High Dispersion Spectrograph \citet[HDS;][]{noguchi02} mounted
 on the Subaru 
Telescope during 2003 to 2010. The wavelength range of 
$\sim 4000 -6800 \AA$ was covered with spectral resolution 
of $R\sim 50000$ for most of the sample stars, while 
the several sample stars (G 64-12, G 64-37, BD+13 2995, G 14-39 
and G 20-15) were observed with $R\sim 90000$. The data reduction 
including bias correction, cosmic-ray removal, 
flat fielding, scattered light subtraction, wavelength calibration, 
and continuum normalization, was performed with standard IRAF routines.    
Details of the observational setting and their reduction  
procedures are described in \citet{ishigaki10} and 
Paper I. The equivalent widths (EWs) of absorption lines 
were measured by fitting Gaussian to each feature.

\subsection{Abundance analysis}

Abundance analyses 
are performed by using an LTE code with  
model atmospheres of \citet{castelli03}, which is widely described 
in \citet{aoki09} and in Paper I. In this subsection, 
we describe additional details on derivation of individual 
elemental abundances. 

\subsubsection{Stellar atmospheric parameters}
\label{sec:params}

We basically adopt the effective temperature ($T_{\rm eff}$), 
surface gravity ($\log g$) and micro-turbulent velocity 
($\xi$) that were estimated and used in Paper I.  
The $T_{\rm eff}$ was estimated by the color $T_{\rm eff}$ relation 
using the calibrations of \citet{casagrande10} for dwarfs and 
\citet{ramirez05} for giants that are based on the infrared flux method. 
As mentioned in Paper I, adopting 
$T_{\rm eff}$ from the color results in a non-negligible slope in the 
iron abundances versus excitation potentials of the \ion{Fe}{1} lines 
for some of the sample stars. 
The two stars HD 171496 and LP 751-19 
show exceptionally large slopes (0.09 and 0.14 dex eV$^{-1}$, respectively) 
compared to the median value of $-0.06$ dex eV$^{-1}$ for the whole sample. 
A possible reason for the peculiar behavior of these stars 
is that the $E(B-V)$ values may be underestimated. 
Both of HD 171496 and LP 751-19 are located at the Galactic latitude, 
$b=-7.7317$ and $-5.2902$, respectively. For these directions, 
the $E(B-V)$ values in \citet{schlafly11} have been reported to 
be $0.38$ and $0.61$, respectively. 
On the other hand, we estimate $E(B-V) = 0.10$ and $0.025$, respectively, 
using the iterative 
algorithm to take into account the finite distance to each star 
(294 and 56 pc, respectively). These estimate may be affected 
by the uncertainty in the distance estimate as well as the 
uncertainty in the distribution of dust near the Galactic disk.  
Since accurate estimates of $E(B-V)$ from interstellar \ion{Na}{1} lines 
is difficult for the spectral resolution of our data, we adopt the $T_{\rm eff}$ 
value of HD 171496 from \citet{alves-brito10}, 
in which the same $T_{\rm eff}$-scale of \citet{ramirez05} was used but with a more sophisticated $E(B-V)$ estimate based on the 
\ion{Na}{1} D line. For LP 751-19, we adopt the value from Paper I. 
As described below, the LP 751-19 shows anomalous abundances for
some elements compared to the behavior of other sample stars, 
which might result from adopting a wrong $T_{\rm eff}$ value.  
The $\log g$ and $\xi$ values of HD 171496 are updated, 
adopting the revised value of $T_{\rm eff}$, 
based on the \ion{Fe}{1}/\ion{Fe}{2} excitation equilibrium and the \ion{Fe}{1} 
abundance-EWs relation.   
  
\subsubsection{Abundances}

We use the EWs of metal absorption lines in the derivation 
of abundances for most of the elements. The measured EWs are given in 
Table \ref{tab:ews}. For Cu and Eu, 
we adopt a spectral synthesis for their abundance estimates.  
The derived abundances are normalized with the solar values 
from \citet{asplund09} to obtain the [X/H] value. The [X/Fe] ratios
are then derived  by normalizing [X/H] 
with [\ion{Fe}{1}/H] or [\ion{Fe}{2}/H] 
for neutral or ionized species, respectively. The derived abundances 
and adopted stellar atmospheric parameters are given in Table \ref{tab:stpm_ab}.

We describe below notes on derivation of individual elemental 
abundances and atomic data. 

{\it Sodium}. 
The sodium abundance is mainly determined from the \ion{Na}{1} lines 
at 5682.6, 6154.2, and 6160.8 {\AA}. We avoid using the \ion{Na}{1} 
resonance lines at 5890/5896 {\AA} since large negative non-LTE 
correction ($\log\epsilon_{\rm NLTE}-\log\epsilon_{\rm LTE}$) 
up to $\sim-0.5$ dex was previously reported for these 
lines \citep{takeda03,andrievsky07}.  On the other hand, for the \ion{Na}{1} 
lines used in the present analysis, 
the non-LTE calculation by \citet{takeda03}
suggests that its correction is not more than $-0.11$ dex for 
their sample of modestly metal-poor ($-1.0<$[Fe/H]$<0.0$) dwarf stars. 
Since the reported amount of correction is not significantly
larger than the errors in the Na abundances in this study, 
we shall consider that correction for the dwarf stars 
in our sample within this metallicity range is negligible. 
The values for the correction may vary 
depending on $T_{\rm eff}$, $\log g$ or metallicities in a complex 
way. However, we assume that the correction is small 
for our whole sample of stars and simply adopt the abundances 
derived from the LTE analysis with no correction.

{\it Scandium}.
The Sc abundances have been determined from EWs of Sc II lines. 
The hyperfine splitting (hfs) was taken into account in the 
abundance derivation, 
adopting the wavelength and the fractional strength of each hyper 
fine component from the 
\citet{kurucz95} database. 
The total $\log gf$ values for each line are normalized to those 
in \citet{ivans06} and \citet{roederer10}. The effect 
of the hfs is very small, 
which is typically $\sim$0.03 dex or less, 
for the present sample.

{\it Vanadium, chromium, nickel, zinc, yttrium, and zirconium}.
The EWs of the \ion{V}{1}, \ion{Cr}{1}, \ion{Cr}{2}, \ion{Ni}{1}, \ion{Zn}{1}, \ion{Y}{2}, and \ion{Zr}{2} lines were 
used for their abundance determination, where $\log gf$ values 
were mainly adopted from \citet{ivans06} and \citet{roederer10}.  
For the \ion{Cr}{1} lines, the $\log gf$ values from 
the recent laboratory measurements of \citet{sobeck07} are 
also included.

{\it Manganese}. For the abundance determination of 
Mn, EWs of \ion{Mn}{1} lines were used. The $\log gf$ 
values of \citet{ivans06}, and \citet{roederer10}, and 
the new measurements of \citet{blackwell-whitehead07} were 
adopted. hfs of these 
lines is taken into account based on the fractional 
strengths of each component in the \citet{kurucz95} database.

{\it Cobalt}. \ion{Co}{1} lines are used for the abundance analysis of Co
taking into account the hyperfine structure for these lines.
The $\log gf$ values and the atomic data for the hyperfine 
structure were taken from \citet{pickering96}.

{\it Copper}. Abundances of Cu have been obtained from the \ion{Cu}{1} line 
at 5105.5 {\AA} for a subset of the sample stars. We have 
employed a spectral synthesis for the abundance estimate, 
since line broadening due to hyperfine and isotopic splitting 
is expected for this line \citep{simmerer03}.  
The line splitting and the fractional strength of each 
hyperfine component were adopted from the \citet{kurucz95} 
database, while the overall $\log gf$ value for 
this line was taken from \citet{fuhr05}. 
The isotopic fractions of the two stable isotopes, 
$^{63}$Cu and $^{65}$Cu, were assumed to be the solar 
system fractions of 69 \%, and 31 \%, 
respectively \citep{simmerer03}.

{\it Strontium}. 
For most of the sample stars, the strong \ion{Sr}{2} lines at 4077.2 {\AA} and 
4215.5 {\AA} are identified. However, these lines are saturated or blended 
especially in the sample stars with [Fe/H]$\gtrsim -1.0$ 
and are not useful for the abundance estimates (a change in 
the Sr abundance makes only small change in the line strength). 
We exclude the sample stars with equivalent widths 
$\log(EW/\lambda)\geq-4.6$ in the following discussions on the Sr abundances.  
The $\log gf$ values for these lines were adopted 
from \citet{fuhr05}.

{\it Barium}.
The Barium abundances have been obtained from EWs of 
the \ion{Ba}{2} lines at 4554.0, 4934.1, 5853.7, and 6141.7 {\AA}. 
 In the abundance calculation, 
hfs and isotopic shifts 
for each of the five stable Ba isotopes 
($^{134}$Ba, $^{135}$Ba, $^{136}$Ba, 
$^{137}$Ba, and $^{138}$Ba) have been 
taken into account. The fractional contribution of 
each hyperfine component and their wavelength shifts 
are basically adopted from \citet{mcwilliam98}, with 
some modifications described below. 

The assumption on the isotopic fraction, which is determined 
by the fractional contribution of the s- and the r-process in synthesizing 
the observed Ba, affects the 
abundance determination, since the odd mass number isotopes 
show hfs while the even mass number isotopes 
do not.  In the present analysis, we adopt the 
isotopic fraction expected for the solar system r-process 
 component from \citet{mcwilliam98}, except for 
the two Ba-rich stars. 
As shown later, these two stars show exceptionally low [Eu/Ba] ratios, which 
indicate a significant contribution of the s-process nucleosynthesis. 
For these two stars, we assume the solar s- and r-process mix
(81\% and 19\%, respectively) predicted by \citet{arlandini99} 
and the isotopic fractions of the s-process component from \citet{anders89}.
The assumption of the solar-system r-process isotopic fraction 
may not be adequate for the sample stars with [Fe/H]$>-1.5$, 
since some contribution from the s-process is expected in this 
metallicity range.  In order to 
reduce the abundance errors due to the uncertainty in the 
isotopic fraction, we exclude the two resonance lines 
at 4554.03 {\AA} and 4934.10 {\AA}, which are particularly 
sensitive to the assumed isotopic fractions
 in the Ba abundance determination 
for the sample stars with [Fe/H]$>-1.5$.
Other two lines in the redder spectral region are relatively 
insensitive to the assumed isotopic fraction and the difference 
in derived Ba abundances when the two assumptions on the
isotopic fraction are made is typically less than 0.02 dex. 
We have also updated the overall $\log gf$ values of each Ba 
line to those recommended by \citet{fuhr05}.

{\it Lanthanum}. The Lanthanum abundances are estimated 
from EWs of \ion{La}{2} lines, taking into account hfs
for these lines.
The $\log gf$ values and the atomic data for the hfs 
were taken from \citet{ivans06}. The \ion{La}{2} lines for which 
hyperfine structure data are not available in \citet{ivans06} are not used 
in the abundance derivation.

{\it Neodymium} and {\it samarium}.
The abundances of neodymium and samarium have been 
obtained from EWs of \ion{Nd}{2} and \ion{Sm}{2} lines, respectively. 
The $\log gf$ values for these lines 
have been adopted from \citet{ivans06}, in which  
the values from recent laboratory measurements were employed. 

\begin{figure}
\includegraphics[angle=270,width=14.0cm]{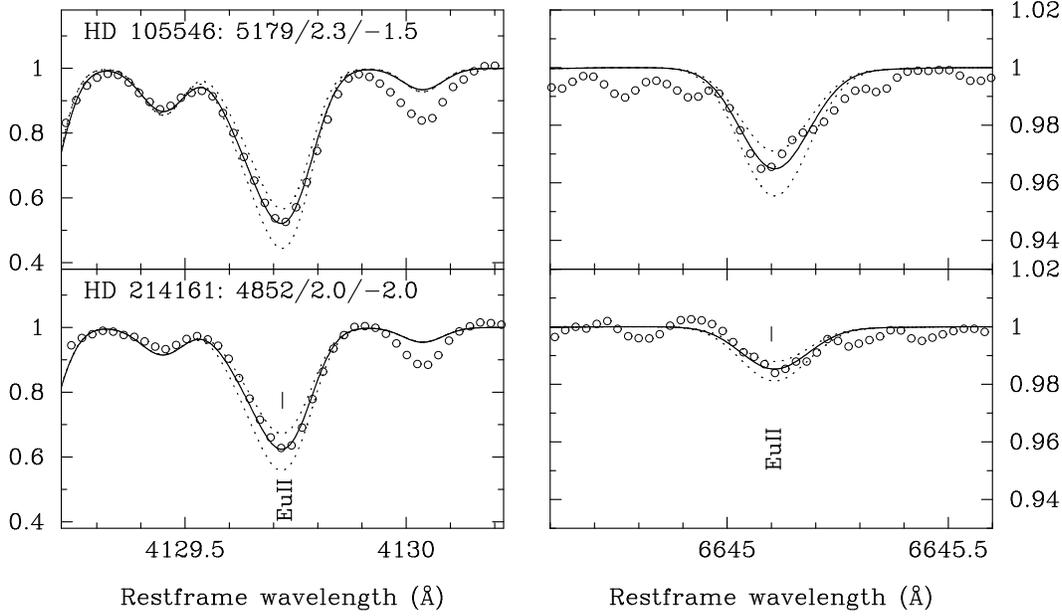}
\caption{Example spectra (circles) of the two sample stars 
for wavelength regions around the 
Eu II 4129.7 {\AA} (left) and 6645.1 {\AA} (right) lines. The adopted atmospheric parameters of these 
stars are indicated on the top of the left panels as 
``$T_{\rm eff}$(K)/$\log g$/[Fe/H]''. The solid lines show
the best-fit synthetic spectra and the dotted lines show the 
spectra for $\Delta \log\epsilon {\rm (Eu)}= \pm 0.1$ dex from 
the best fit values.}
\label{fig:euspec}
\end{figure}

{\it Europium}.
Europium abundances have been determined 
from the spectral synthesis of \ion{Eu}{2} 4129.7 and/or 6645.1 {\AA} lines.  
An example of the fitted spectrum is shown in Figure \ref{fig:euspec}.
Abundance fraction of two naturally occurring isotopes, $^{151}$Eu and 
$^{153}$Eu is assumed to be 
$^{151}$Eu$\equiv$$^{151}$Eu/($^{151}$Eu+$^{153}$Eu)$=0.5$, which is 
roughly consistent with the r-process component of the 
Solar-system meteorite abundance \citep{anders89}. 
Hyperfine and isotopic structures for these lines 
were calculated based on the data listed in \citet{lawler01}.
The oscillator strengths were also taken from \citet{lawler01}. 
The relative strengths of the transitions were computed with  
a standard manner as described in \citet{lawler01}. 
For lines of other species within a few {\AA} of each \ion{Eu}{2} line, 
we have adopted the $\log gf$ values from the current 
version of the \citet{kurucz95} 
database. The \ion{Eu}{2} 6645 {\AA} 
line is probably contaminated by an \ion{Si}{1} line at 6645.21 {\AA} 
in the adopted line list. 
The $\log gf$ value of this line was slightly modified so that 
the calculated synthetic spectrum for the Sun best reproduces the 
observed solar spectra in the wavelength range surrounding 
this line. \citet{mashonkina12} 
reported that non-LTE correction for the \ion{Eu}{2} 4129 {\AA} line ranges 
from 0.05-0.12 dex and is generally larger 
for lower gravity stars (i.e., giants). In order to avoid 
systematic errors due to the non-LTE effect, we 
present the results considering dwarfs and giants
separately when comparing the 
Eu abundances among the subsamples (see Section \ref{sec:eu}).

\subsubsection{Abundance errors}

Errors in the abundances are computed by taking into account 
line-to-line scatter in derived abundances and uncertainty in 
the adopted atmospheric parameters as in Paper I. 
The line-to-line scatter in the abundances from individual lines is 
typically smaller than 0.10 dex. Errors in the mean of the 
abundances due to the scatter are calculated as the line-to-line 
scatter divided by a 
square root of the number of the lines used to compute the mean. 
When only one line is used to estimate the abundances, we assume that the 
error is equal to the line-to-line scatter in \ion{Fe}{1} lines, which are 
typically the most numerous.  
Abundance errors due to the uncertainty in the adopted 
$T_{\rm eff}$, $\log g$, and $\xi$ values are examined 
by changing these parameters by $\pm 100$ K, $\pm 0.3$ dex, and 
$\pm 0.3$ km s$^{-1}$, respectively, in the abundance estimates. 
The final errors are obtained by summing these contributions 
in quadrature and are listed in Table \ref{tab:stpm_ab}.

\subsection{Comparison with other studies}

\begin{figure}
\includegraphics[height=11.0cm]{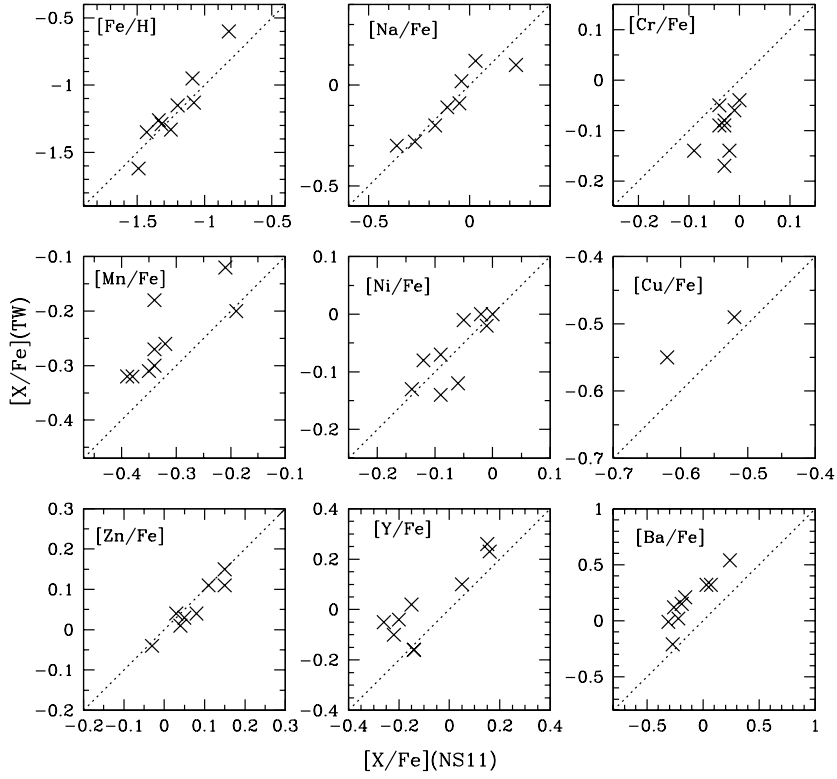}
\caption{Comparison of the derived [Fe/H], [Na/Fe], [Cr/Fe], [Mn/Fe], [Ni/Fe], [Cu/Fe], [Zn/Fe], [Y/Fe] and [Ba/Fe] abundance ratios with those from \citet{nissen10,nissen11}}
\label{fig:comp_nissen}
\end{figure}

Figure \ref{fig:comp_nissen} shows the 
comparison of derived abundance 
ratios with those from \citet{nissen10, nissen11} for the nine stars 
analysed in common  (G 112-43, G 53-41, G 125-13, G 20-15, G 176-53, 
G 188-21, HD 111980, HD 105004, and HD 193901).  The abundance results 
of the two studies are also summarized in Table \ref{tab:comp_nissen}.
For [Na/Fe], [Ni/Fe], and [Zn/Fe], the derived abundances show 
an excellent agreement within 0.01 dex with the rms scatter for the difference 
of $\leq$0.07 dex. The [Cr/Fe], [Mn/Fe], [Cu/Fe], and [Y/Fe] in 
the two studies marginally agree within the mean differences of 
at most 0.09 dex. 
The derived [Ba/Fe] ratios tend to be larger 
in this work than in \citet{nissen11} by 0.28 dex on average with 
scatter of 0.10 dex.  
The large difference is partly attributed to the difference in 
the adopted microturbulent velocity ($\xi$). As shown in 
Paper I, the present study has adopted systematically 
lower $\xi$ than that in \citet{nissen11}, which results in 
the larger Ba abundances. 
Another  possible cause for the discrepancy is the difference in 
the adopted damping constant, for which the \citet{unsold55} approximation 
to the Van der Waals constant, enhanced by a factor of 2.2 was 
employed in this study.

\begin{figure}
\includegraphics[height=11.0cm]{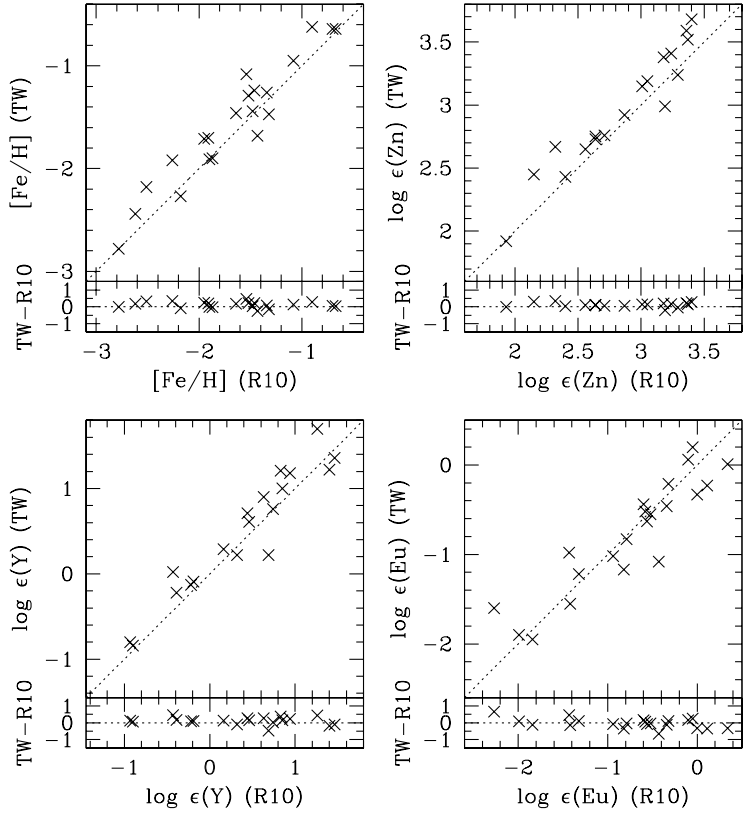}
\caption{Comparison of the derived [Fe/H], $\log \epsilon ({\rm Zn})$, $\log \epsilon ({\rm Y})$, and $\log \epsilon ({\rm Eu})$ abundances with those from 
\citet{roederer10}}
\label{fig:comp_roederer}
\end{figure}

Figure \ref{fig:comp_roederer} shows comparisons of 
derived [Fe/H], $\log \epsilon ({\rm Zn})$, $\log \epsilon ({\rm Y})$, and $\log \epsilon ({\rm Eu})$ abundances with those from 
\citet{roederer10} and \citet{simmerer04} for the 21 stars studied in common. 
The abundance results from these studies and this work are summarized in 
Table \ref{tab:comp_roederer}.
For [Fe/H], $\log\epsilon ({\rm Zn})$, and $\log\epsilon ({\rm Y})$, 
our abundances are systematically higher, where   
means of the difference (scatter) are $\Delta$[Fe/H](TW-R10)=0.12 (0.17), 
$\Delta\log\epsilon ({\rm Zn})$(TW-R10)=0.12 (0.12), 
and $\Delta\log\epsilon ({\rm Y})$(TW-R10)=0.10 (0.21) dex.
These offsets could partly be attributed to the difference in the adopted 
$T_{\rm eff}$, which is higher in this study by $\sim 60$ K on average than 
those of \citet{roederer10} taken from \citet{simmerer04}. 
The mean of the difference is smaller 
for $\log\epsilon ({\rm Eu})$ (-0.01 dex)
but with a larger scatter of 0.28 dex. The larger 
scatter for $\log\epsilon ({\rm Y})$ 
and $\log\epsilon ({\rm Eu})$ is partly due to a few  
stars for which large discrepancy is found. 
For the sample star G 63-46 with $\Delta\log\epsilon ({\rm Y})$(TW-R10)$=0.44$ 
dex, the difference is likely attributed to $\sim 160$ K difference in 
the adopted $T_{\rm eff}$. For another star HD 128279 with  
 $\Delta\log\epsilon ({\rm Eu})$(TW-R10)$=0.67$ dex, only single
\ion{Eu}{2} line, which is close to the detection limit, 
is used for the abundance estimate in the present study. 
Thus, the uncertainty in the synthetic spectral fitting 
may be mainly responsible for the discrepancy.

\subsection{[X/Fe]-$T_{\rm eff}$ correlation}
\label{sec:abu_teff}

The sample stars in the present study have various $T_{\rm eff}$ 
values in the range $4000-6900$ K. 
In this subsection, we examine the [X/Fe]-$T_{\rm eff}$ correlation 
among the sample stars and examine the extent to which such a 
correlation might affect the abundance comparison 
between the thick disk, inner halo, and outer halo subsamples. 

\begin{figure}
\includegraphics[height=13.0cm]{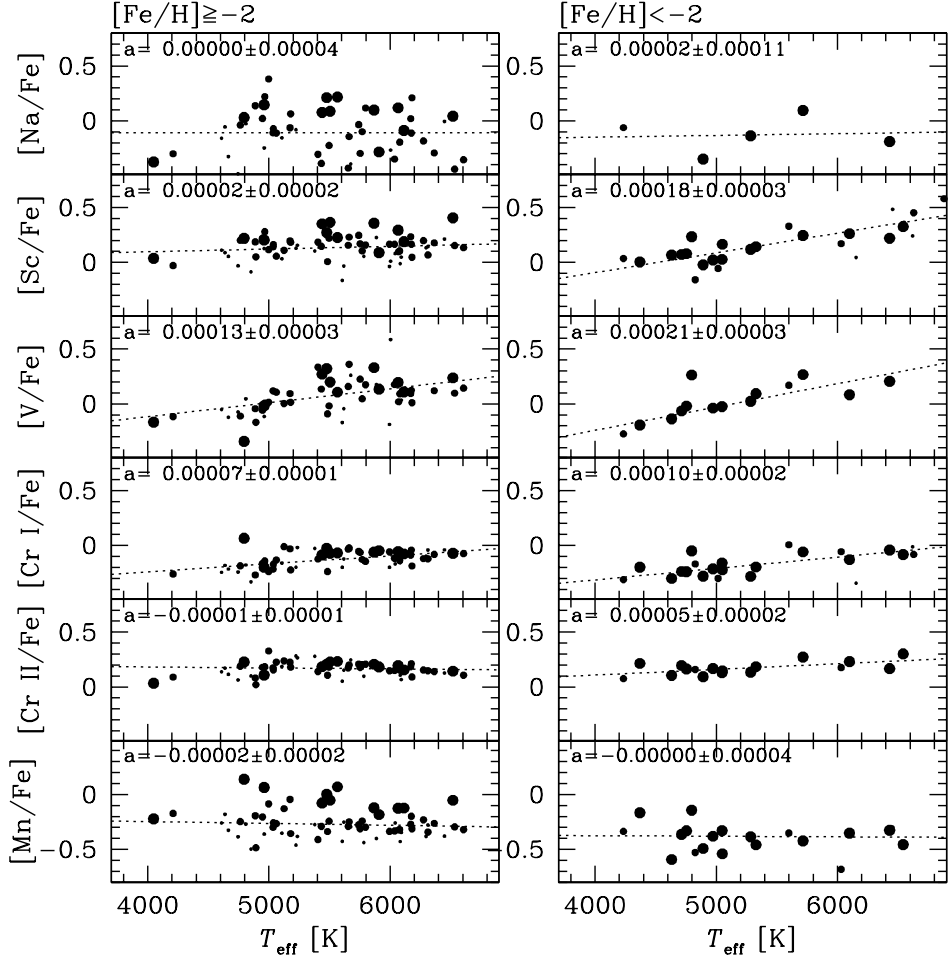}
\caption{Abundance ratios of 
Na, Sc, V, \ion{Cr}{1}, \ion{Cr}{2} and Mn plotted against the adopted 
$T_{\rm eff}$ values for the sample stars with 
the metallicity [Fe/H]$\geq -2$ (left) and [Fe/H]$<-2$ (right). The size of the symbols corresponds to metallicity; in the left (right) panel, small:$-2.0\leq$[Fe/H]$<-1.5$ ([Fe/H]$<-3.0$), medium:$-1.5\leq$[Fe/H]$<-1.0$ ($-3.0\leq$[Fe/H]$<-2.5$), and large:$-1.0\leq$[Fe/H] ($-2.5\leq$[Fe/H]$<-2.0$). A dotted line in 
each panel shows the result of a least-squares fit to a straight line [X/Fe]$=b+aT_{\rm eff}$. 
The slope $a$ of the fit is indicated in each panel. 
}
\label{fig:fepnc1_teff}
\end{figure}

\begin{figure}
\includegraphics[height=13.0cm]{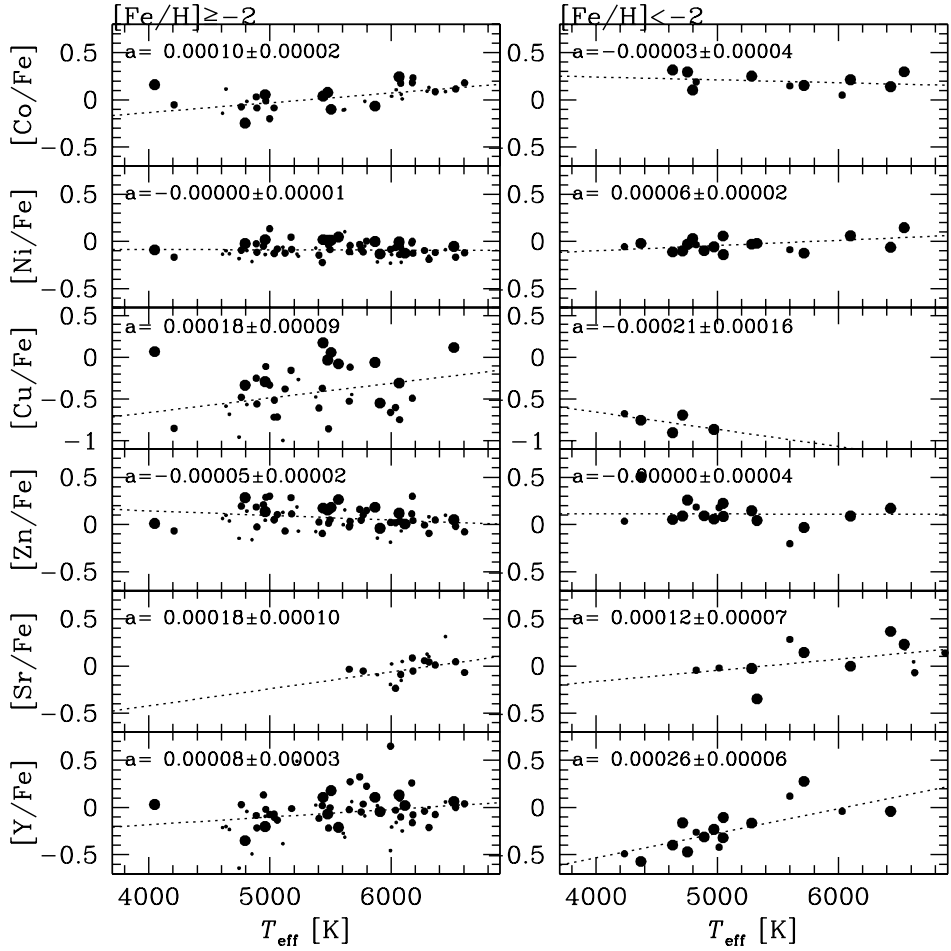}
\caption{Same as Fig \ref{fig:fepnc1_teff} but for Co, Ni, Cu, Zn, Sr, and Y.}
\label{fig:fepnc2_teff}
\end{figure}

\begin{figure}
\includegraphics[height=13.0cm]{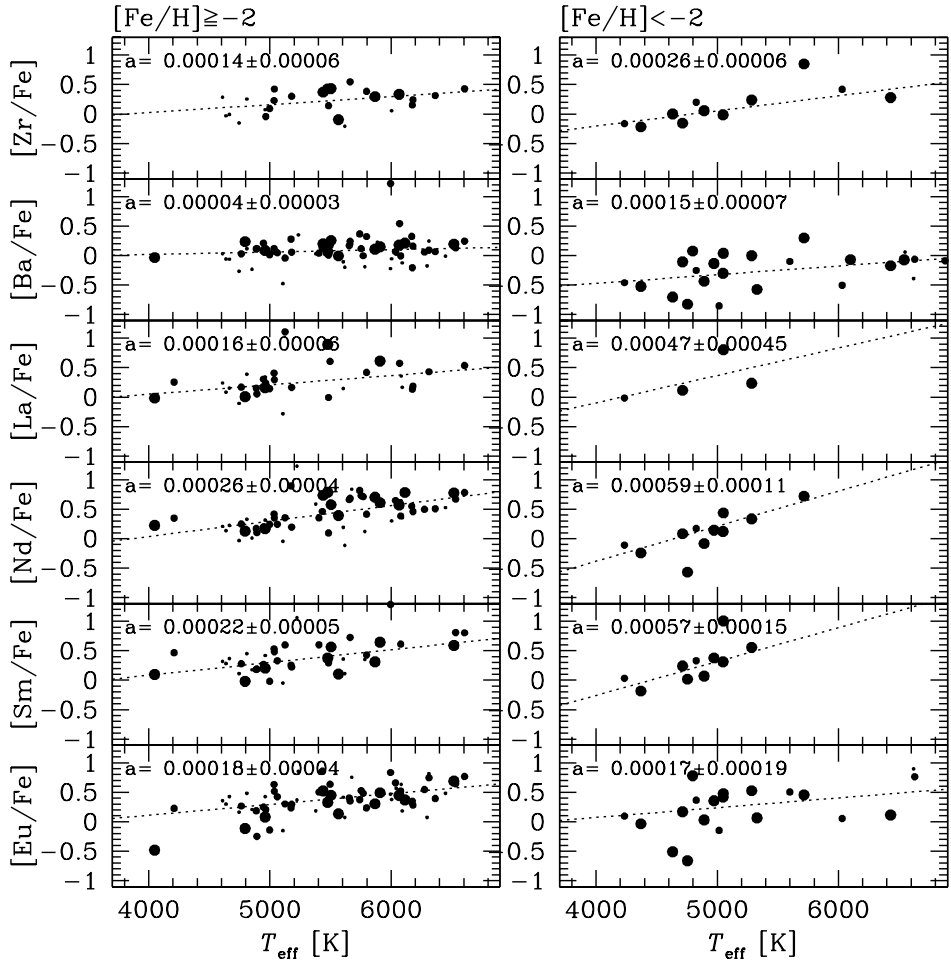}
\caption{Same as Fig \ref{fig:fepnc1_teff} but for Zr, Ba, La, Nd, Sm., and Eu.}
\label{fig:fepnc3_teff}
\end{figure}

Figures \ref{fig:fepnc1_teff}-\ref{fig:fepnc3_teff} show the [X/Fe]
plotted against $T_{\rm eff}$. 
The left and right rows of each figure show the plots for 
[Fe/H]$\geq -2$ and $<-2$, respectively. 
The three sizes of symbols represent the three metallicity intervals with 
the larger symbols corresponding to higher metallicities (see the 
caption of Figure \ref{fig:fepnc1_teff}). 
A slope of the linear regression line, which is calculated 
by a two-sigma clipping algorithm, is indicated in the top 
of each panel. 

It can be seen from the figures that some of the elements show 
a slope larger than 3$\sigma$ in the 
[X/Fe]-$T_{\rm eff}$ plane in one or both 
metallicity range(s).
In [Fe/H]$\geq -2$ (the left columns of  Figures \ref{fig:fepnc1_teff}-\ref{fig:fepnc3_teff}), a significant [X/Fe]-$T_{\rm eff}$ 
correlation can be recognized for V, \ion{Cr}{1}, Co, Nd, Sm, and Eu. 
For V, Cr, and Co, the slopes are $\le 0.13$ dex/1000K, which is 
comparable to the observational errors, while for Nd, Sm, and Eu, 
the slopes exceed $\ge 0.18$ dex/1000 K, which may affect 
the abundance comparison between the subsamples. 
In [Fe/H]$<-2$ (right columns of Figures \ref{fig:fepnc1_teff}-
\ref{fig:fepnc3_teff}),  the 
abundance ratios of Sc, V, \ion{Cr}{1}, Y, Zr, La, Nd, and Sm 
show a slope of $>3\sigma$ with $T_{\rm eff}$. 
 Particularly large slopes for [Nd/Fe] and 
[Sm/Fe] versus $T_{\rm eff}$ plots are partly attributed to 
paucity of data points 
in the range $T_{\rm eff}>5000$ K, in which Nd and Sm abundances 
are below detection limits for many of the sample stars.
Since the upper limits for some of the sample stars in this 
temperature range are below [Nd/Fe], [Sm/Fe]$\sim$0.5 dex, the 
apparent extreme slopes are likely artificial.

The abundance-$T_{\rm eff}$ correlations as indicated above or 
the discrepancy in the derived abundances between dwarf and giant stars have 
been reported in previous studies. 
\citet{bonifacio09} compare abundances in the sample of dwarfs 
and giants in the metallicity range of $-4\lesssim$[Fe/H]$\lesssim -2$. 
They reported that the dwarf versus giant discrepancy presents for 
many elements. In particular, the Sc, Cr, Mn, Zn, and Co abundances 
were found to be higher in dwarf stars than in giant stars. 
This effect is also seen in our sample for Sc and \ion{Cr}{1} in a 
similar metallicity range. On the other hand, our sample does not 
show significant dwarf/giant discrepancy for Mn, Zn, 
and Co. 

The exact reason for the discrepancy is currently unclear. 
\citet{bonifacio09} suspect that granulation in the stellar atmospheres
 (three-dimensional effects) and/or departure from LTE might be responsible 
for the discrepancy observed in some elements. 
Since the magnitudes and direction 
of these effects may be different among different species and lines 
used in the analysis, we do not correct 
[X/Fe] values to vanish the [X/Fe]-$T_{\rm eff}$ slopes in the following 
analysis. 
Instead, for the elements with the large slopes, 
we separately treat dwarfs and giants in the 
abundance comparisons among the three subsamples.

\section{Results}
\label{sec:results}

\subsection{Distribution of the sample stars in [X/Fe]-[Fe/H] planes}

\begin{figure}
\includegraphics[width=14.0cm]{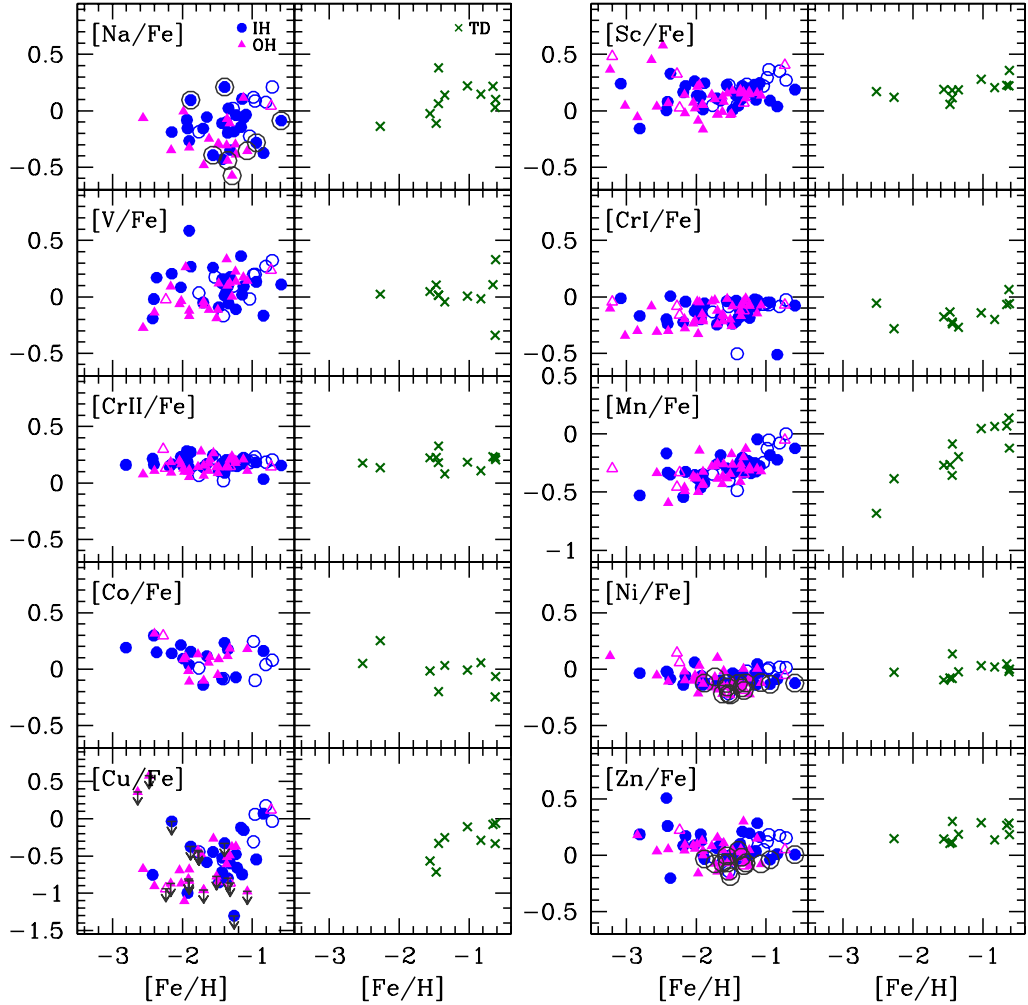}
\caption{Abundance ratios for Na, Sc, V, \ion{Cr}{1}, \ion{Cr}{2}, 
Mn, Co, Ni, Cu, and Zn plotted against [Fe/H]. The crosses, 
filled circles, and filled triangles 
indicate the sample stars with $P_{\rm TD}>0.9$ (the thick disk stars), 
$P_{\rm IH}>0.9$ (the inner halo stars) and $P_{\rm OH}>0.9$ 
(the outer halo stars), respectively. Open circles show the 
stars whose kinematics are intermediate between the thick disk 
and the inner halo ( $P_{\rm TD}, P_{\rm IH}\leq 0.9$ and $P_{\rm TD}$, $P_{\rm IH}\geq P_{\rm OH}$ ), while open triangles indicate the stars 
whose kinematics are 
intermediate between the inner and the outer halo 
($P_{\rm IH}$, $P_{\rm OH}\leq 0.9$ and $P_{\rm IH}$, $P_{\rm OH}\geq P_{\rm TD}$). Symbols marked with a gray open circle in the [Na/Fe], [Ni/Fe] and [Zn/Fe] 
panels represent the sample stars with [Mg/Fe]$<0.1$.}
\label{fig:fepnc1_mem}
\end{figure}

Figures \ref{fig:fepnc1_mem} and \ref{fig:fepnc2_mem} 
show the abundance ratios ([X/Fe]) 
plotted against [Fe/H] for the thick disk stars (crosses), the inner halo stars 
(filled circles), the outer halo stars (filled triangles), and their 
intermediate populations (thick/inner halo: open circles, inner/outer halos
: open triangles). 
For particularly interesting elements, namely, Na, Ni, Zn, and Eu, 
we additionally discuss the behavior 
 of low-[Mg/Fe] stars with [Mg/Fe]$<0.1$, which are 
analogous to the low-$\alpha$ 
 stars in \citet{nissen10} for the purpose of examining the consistency 
of the present results with those of \citet{nissen10}. 
The low-[Mg/Fe] stars are marked with gray circles in the corresponding 
panels in Figures \ref{fig:fepnc1_mem} and \ref{fig:eu_mg}. 
Table \ref{tab:mean_dev} summarizes means ($\mu$) and scatters ($\sigma$) of 
the abundance ratios for the three subsamples (in the second 
column, "TD", "IH", and "OH" for the thick disk, inner halo, 
and outer halo subsamples, respectively). This table also 
includes the means and scatters of the abundance ratios taking into account 
dwarfs ($\mu_{\rm d}$) or giants ($\mu_{\rm g}$) alone. 
The anomalous star, LP 751-19, (see Section \ref{sec:params}), 
is excluded in the calculation of the means and scatters.

\subsubsection{Sodium, Scandium and Vanadium}
\label{sec:result_sodium}

The thick disk, inner halo, and 
outer halo subsamples show different trends and scatters 
in the [Na/Fe]-[Fe/H] diagram.  
First, the thick disk stars are modestly 
enhanced in [Na/Fe] ratios at metallicities 
[Fe/H]$>-1.5$ with a mean abundance of [Na/Fe]$=0.10$ dex, 
while most of the inner and outer halo stars show 
lower ratios at similar metallicities.  
Second, scatter in the [Na/Fe] ratios in this [Fe/H] range is 
relatively small (0.11 dex) for the thick disk stars, which is 
comparable to the observational error, 
while the inner and outer halo stars show larger scatter ($\geq 0.17$ dex). 
Third, the mean [Na/Fe] ratio for the outer halo stars ($-0.28\pm 0.06$ dex) 
is lower than that of the inner halo stars ($-0.12\pm 0.04$). There 
are also abundance differences among the three subsamples 
when only dwarfs or giants are taken into account 
 (see Table \ref{tab:mean_dev}). 
Furthermore, among the inner halo and outer halo stars, those with lower
[Mg/Fe] ratios (gray circled symbols) tend to show 
lower [Na/Fe] ratios in a given metallicity range with a possible 
exception of the two stars having [Na/Fe]$>0.0$. 

The observed trends of [Na/Fe] ratios with [Fe/H] for 
our three subsamples are in agreement with previous studies. 
\citet{reddy06} reported modestly enhanced 
[Na/Fe] ratios and decreasing [Na/Fe] with decreasing [Fe/H] 
for their sample of thick disk stars with [Fe/H]$<-0.6$, 
which is similar to the trend seen in the plot in Figure \ref{fig:fepnc1_mem}. 
The non-LTE (re)analysis of sodium abundances by \citet{takeda03} 
suggests that [Na/Fe] ratios in their sample of thick disk 
stars are near-solar, which is roughly consistent with 
the present result if the suggested non-LTE correction up to 
$\sim -1.0$ dex is applied to our sample. 
 
For the inner and outer halo stars, our results 
suggest that the [Na/Fe] ratios are likely
correlated with both kinematics and [$\alpha$/Fe], which qualitatively 
supports the previous findings by \citet{stephens02},\citet{fulbright02}, and 
\citet{nissen10,nissen11}. 
As an example, \citet{fulbright02} reported that a fraction of 
Na-poor stars defined as [Na/Fe]$<-0.36$ in his sample increases 
 for stars with a large apocentric distance ($R_{\rm apo}>20$ 
kpc). The suggested dependence is similar to 
the present result in that the outer halo 
subsample, which includes stars with $R_{\rm apo}> 15$ kpc,
tends to show relatively low [Na/Fe] ratios than the inner halo 
stars. We will discuss the likely [Na/Fe]-kinematics 
correlations in Section \ref{sec:abu_kin}.
\citet{nissen10} reported the 
distinct [Na/Fe] ratios for the low- and high-$\alpha$ stars 
in the metallicity range of $-1.6<$[Fe/H]$<-0.4$, for which 
a similar trend is apparent in Figure \ref{fig:fepnc1_mem}.
To summarize, the inner and outer halo stars, at least 
in part, show the lower [Na/Fe] ratios 
than the thick disk stars and the lowest 
[Na/Fe] stars in our sample tend to have 
outer halo kinematics and/or low-[Mg/Fe] ratios.

The observed difference in the [Na/Fe] ratios
among the three subsamples suggests that progenitors of the  
thick disk and the inner and outer halos experienced 
largely different chemical enrichment histories. 
Sodium is mainly synthesized during hydrostatic carbon
 burning in the massive stars \citep{woosley95} and 
its yield is known to be dependent on mass and metallicity 
of the progenitor star \citep{kobayashi06}. Therefore, 
the IMF, typical metallicities,  
and/or the relative contribution of Fe from Type Ia SNe
 may be responsible for determining [Na/Fe] abundance ratios 
of the progenitor systems. 
One possible interpretation for the origin of the observed 
lower [Na/Fe] stars in our inner and outer halo subsamples 
is that higher mass stars were deficient in their 
progenitors compared to those of the thick disk. 
Alternatively, metals ejected from massive stars 
more easily escaped in the progenitor of the inner and outer halos 
than those of the thick disk.
Star formation rate is another factor that could affect [Na/Fe] ratios 
since it determines the relative contribution of Na predominantly 
from Type II SNe to Fe from Type Ia SNe. 
In order to examine which of the above factors is the most 
important for explaining
the observed [Na/Fe] difference among the three subsamples, 
a chemical evolution modeling which takes into account 
the difference in star formation environment during the formation 
of the three components is necessary.

The [Sc/Fe] ratios for the thick disk, inner halo, and 
 outer halo subsamples are all enhanced in the metallicity
range [Fe/H]$>-1.5$ with a modest 
decreasing trend toward lower [Fe/H] with 
small scatter ($<0.10$ dex). 
A significant difference in this trend among the 
three subsamples is not found. 
In [Fe/H] below $\sim -1.5$, the trend appears to be
 flattened at [Sc/Fe]$\sim 0.10$ dex
 showing a much larger scatter ($>0.25$ dex; Table \ref{tab:mean_dev}).
The observed enhanced [Sc/Fe] ratios
 are generally consistent with those found 
in previous studies for stars with [Fe/H]$>-1.5$ \citep{prochaska00, 
reddy06,zhao90,nissen00,cayrel04}.

The [V/Fe] ratios for the three subsamples are 
nearly flat at the solar value in the metallicity below 
[Fe/H]$\sim -1.0$ without a significant difference among 
the subsamples. 
The near-solar value for the halo stars is in agreement 
with previous studies \citep{gratton91,lai08}. 
The enhancement in the [V/Fe] ratio
in [Fe/H]$>-1.0$ is reported by 
\citet{prochaska00} and \citet{reddy06} for the thick disk 
stars, while a modest enhancement (0.02-0.13 dex) is also seen 
in our sample. 

In the theoretical calculation of \citet{woosley95}, 
both Sc and V are produced through explosive oxygen, silicon 
and neon burning in massive stars and thus their 
yield is sensitive to various parameters of
SN explosions. The similarity in the observed 
trend in [Sc/Fe] and [V/Fe] among the three subsamples may indicate that 
the astrophysical sites for the production of these elements 
are largely common among the different Galactic populations.

\subsubsection{Chromium}

For chromium, we first note that the 
abundances derived from the neutral and 
ionized species ([\ion{Cr}{1}/Fe] and [\ion{Cr}{2}/Fe], respectively)  
are systematically different, as can be seen in the 
plot in Figure \ref{fig:fepnc1_mem}.
The [\ion{Cr}{1}/Fe] ratios are subsolar for the whole [Fe/H] range and 
slightly decrease toward lower metallicity, while the [\ion{Cr}{2}/Fe] ratios  
are supersolar without any trends with [Fe/H].   
The observed discrepancy in the Cr abundance ratios 
between those derived from neutral and ionized species 
has previously been reported in the literature 
\citep{gratton91,lai08,bonifacio09}. 
The reason for this discrepancy and which species is the more reliable 
indicator of the true Cr abundance are not clear. 
\citet{gratton91} suggested that the neutral 
species is affected by overionization and thus use of \ion{Cr}{1} lines 
would underestimate the overall Cr abundance. 

As far as the [\ion{Cr}{2}/Fe] ratio is concerned, which is presumably 
less affected by 
the overionization, the abundance ratios show a negligible 
scatter of 0.06 dex at most for all of the thick disk, inner halo 
and outer halo subsamples over the whole [Fe/H] range. This result 
is consistent with that of a more precise analysis of \citet{cayrel04}, 
which reported the $0.05$ dex scatter over the metallicity $-4<$[Fe/H]$<-2$. 
The present results further confirm the small cosmic scatter 
in the [Cr/Fe] ratios in the solar neighborhood stars 
independent of their kinematics and metallicity.

\subsubsection{Manganese, Nickel and Zinc}

For all of the three subsamples, the [Mn/Fe] ratios 
show an increasing trend with increasing [Fe/H]
in a range [Fe/H]$>-2$ with 
modest scatter ($\leq 0.09$ dex). In a more metal-poor range, the 
trend appears to be flattened with larger scatter. 
The increasing [Mn/Fe] trend is consistent with previous studies for 
the thick disk and halo stars 
in the range $-1.5<$[Fe/H]$<0.0$ \citep{prochaska00,nissen00,reddy06}.  
Systematic differences between the 
thick disk, inner halo, and outer halo stars are not clearly 
seen. 

Manganese is thought to be 
mostly produced by explosive silicon burning in massive 
stars in their outer incomplete Si-burning layers \citep{umeda05} 
and in Type Ia SNe \citep{iwamoto99}. 
In metallicities below [Fe/H]$\sim -1$, the low [Mn/Fe] ratios 
are mainly determined by the yields from Type II SNe of massive stars 
\citep{tsujimoto98}. In [Fe/H]$\gtrsim -1$, the increasing [Mn/Fe] 
with increasing [Fe/H] is interpreted as an onset of 
contribution from Type Ia SNe \citep{kobayashi06}. 
It was also suggested that the dependence of Mn yields of Type 
Ia SNe on metallicity may contribute to the increase of [Mn/Fe] ratios 
with [Fe/H] \citep{cescutti08}. 
The observed increase in the [Mn/Fe] ratios for the three subsamples, 
therefore, may indicate that 
Type Ia SNe have played some role for chemical evolution in the 
progenitors of the thick disk, inner halo, and outer halo components. 
This interpretation favors the idea that the formation timescales for these 
progenitors were modestly 
longer than those of the Type Ia SNe. The timescale for 
the chemical enrichment via Type Ia SNe is poorly 
constrained due to the uncertainties in physical mechanisms 
that give rise to the explosion \citep[e.g.][]{maoz10}.
Analyses of SN rates in galaxies and galaxy clusters suggest 
that the SN rate as a function of the delay time from major star formation 
to SN explosions is higher within a few Gyr and decrease 
toward longer time delays \citep{totani08,maoz10}.
Such estimates for the SN delay time distributions as well as 
constraints on SN Mn yields are essential to interpret
the [Mn/Fe] ratios for each of the three subsamples.

For the [Ni/Fe] ratios, a modest difference between the thick disk 
and inner/outer halo subsamples can be recognized. 
As indicated in Table \ref{tab:mean_dev}, 
the thick disk stars show near solar [Ni/Fe] ratios while the inner halo 
and outer halo stars show $\sim 0.10$ dex lower [Ni/Fe] ratios 
in [Fe/H]$>-2.5$. The difference between the two halo subsamples 
is not very clear. All of the subsamples show relatively small 
scatter ($\leq 0.09$ dex) of [Ni/Fe] ratios about the mean value. 
\citet{nissen10} reported the underabandance of Ni in low-$\alpha$ 
stars. Although the present study does not have very high precision to 
convincingly confirm this argument, the low-[Mg/Fe] stars (gray circled symbols) tend to show lower 
[Ni/Fe] than the other stars (predominantly the thick disk stars and 
thick disk/inner halo intermediate stars) in [Fe/H]$\gtrsim -1.2$. 

Nickel isotopes are produced in both deep layers 
of massive stars and in Type Ia SNe \citep{woosley95,timmes95}. The 
observed difference in [Ni/Fe] among the thick disk and the inner/outer 
halo stars at [Fe/H]$>-2.5$ may indicate that the relative 
contribution of massive stars and Type Ia SNe to the Ni production 
is different among the progenitors of these subsamples. 

The [Zn/Fe] ratios show an interesting difference between the subsamples 
as can be seen in the bottom left panel of Figure \ref{fig:fepnc1_mem}. 
The thick disk stars show super-solar [Zn/Fe] values with relatively small 
scatter over all metallicities. The inner halo stars show slightly 
lower [Zn/Fe] particularly in [Fe/H]$>-1.0$ with a modest scatter. 
The outer halo stars show lower [Zn/Fe] than the other two subsamples 
in the intermediate metallicity range with a scatter similar 
to the inner halo stars. One outer halo star G 112-43, which constitutes a
 common-proper-motion system with G 112-44, shows an exceptionally
 high [Zn/Fe] abundance compared to the other outer halo stars. 
Such peculiar abundance for this star is previously noted by 
\citet{nissen10, nissen11}. 

\citet{nissen11} reported that in $-1.6<$[Fe/H]$<-0.4$, the thick disk 
stars and the high-$\alpha$ halo stars show constantly high-[Zn/Fe] ratios 
while those of the low-$\alpha$ stars show a mildly decreasing trend 
with [Fe/H]. 
In this metallicity, a similar difference can be recognized in our sample:  
the [Zn/Fe] ratios of low-[Mg/Fe] stars (gray circled symbols)
 are lower than other stars on average. 
Additionally, the difference seems to continue toward a lower metallicity 
of [Fe/H]$\sim -2.0$ in the present sample. 
In the much lower metallicity range ([Fe/H]$<-2$), both the inner halo and 
the outer halo stars show supersolar [Zn/Fe] ratios in agreement with previous 
studies \citep{cayrel04,takeda05,lai08}, except for the low-[Mg/Fe] stars.

At low metallicities ([Fe/H]$\lesssim-2.5$),  Zn is thought to
 be produced in the deep complete Si burning region in massive 
 Type II SNe \citep{umeda05,kobayashi06}. 
\citet{umeda05} suggest that the ejection of Zn is 
more enhanced in higher energy SNe, which is called hypernovae. 
Zinc is also thought to be synthesized through 
a weak s-process component 
in massive stars \citep{timmes95}. 
The observed supersolar [Zn/Fe] in our sample, therefore, could 
be a signature of the hypernovae and/or 
neutron-capture process in massive stars. 
 In the higher metallicity, 
the contribution from the s-process in low-mass asymptotic giant branch 
(AGB) stars 
and Fe production through Type Ia SNe could affect the 
observed [Zn/Fe] \citep{timmes95}. 
The [Zn/Fe] differences among the three subsamples 
predominantly seen 
in [Fe/H]$>-2.0$ indicate different contributions 
of Type Ia and AGB products in the progenitor interstellar medium (ISM) of 
these populations. Additionally, difference in 
mass and/or metallicity of stars responsible for the 
Zn production may also affect the observed abundance 
differences. Deeper understanding of the [Zn/Fe] 
among different populations 
require more robust estimates of Zn yields and their production timescales 
in various astrophysical sites.

\subsubsection{Cobalt and Copper}

The [Co/Fe] ratios of the thick disk, inner halo and outer 
halo stars all show similar behavior against [Fe/H], namely, 
the [Co/Fe] is close to the solar value in [Fe/H]$>-2.0$, while it
is $\sim 0.2$ dex in lower metallicities. The scatter of 
0.09-0.19 dex may partly be attributed to the 
observational error, since only one line, which is particularly strong 
in giant stars, is used for the abundance estimate in 
most of the sample stars. 
 The enhanced [Co/Fe] ratios in [Fe/H]$<-2.0$ generally agree with 
the results of \citet{cayrel04} and \citet{lai08}.

Cobalt is produced in both Type II and Type Ia SNe 
\citep{timmes95,kobayashi06}. 
A chemical evolution model of \citet{timmes95} suggests that the
trend in [Co/Fe] with respect to [Fe/H] is determined by dependence of 
Co ejection on mass/metallicity of massive stars and by the production of 
Fe through Type Ia SNe. The observed similarity in [Co/Fe] 
among the the thick disk, inner halo and outer halo subsamples 
is in contrast to the [Zn/Fe], for which abundance differences 
among the three subsamples are identified. 
This result suggests that, unlike [Zn/Fe], 
[Co/Fe] ratios are relatively insensitive to star formation/
chemical enrichment histories in various progenitor systems.

The [Cu/Fe] ratios are plotted against [Fe/H] in the 
bottom right panel of Figure \ref{fig:fepnc1_mem}. The downward arrows 
are overlaid for the sample stars for which only 
an upper limit has been obtained.
The thick disk stars show near to subsolar [Cu/Fe] ratios 
in the range $-0.5$ to 0.0.
Both of the inner halo and outer halo stars 
show lower [Cu/Fe] ($\sim -0.8$) at metallicities 
up to [Fe/H]$\sim -1.0$ with much larger scatter. The lower 
[Cu/Fe] for the halo stars than the thick disk stars is 
consistent with the results of  \citet{mishenina02} and \citet{reddy06}. 
An increasing [Cu/Fe] trend 
with [Fe/H] is modestly seen in the thick disk 
and the inner halo subsamples, while the trend is not 
clear in the outer halo subsample.

A large fraction of Cu is thought to be produced 
 in a neutron-capture process in massive stars during 
their convective core He-burning and the shell C-burning 
\citep[e.g.][]{pignatari10}. 
The neutron source in these stars is mostly provided by 
the $^{22}$Ne($\alpha$,n)$^{25}$Mg reaction, where $^{22}$Ne is 
produced from CNO isotopes. Thus, in this scenario, the production of 
Cu depends on the initial CNO composition of the 
progenitor star. 
This production channel of Cu may explain the 
modest increasing [Cu/Fe] trend with [Fe/H] for 
the thick disk and the inner halo stars. 
If this production channel 
is dominant at [Fe/H]$>-1.5$, the lower [Cu/Fe] ratios for 
some inner/outer halo stars may require extra enrichment of Fe presumably 
from Type Ia SNe.

\subsubsection{Light neutron-capture elements: Sr, Y, and Zr}

\begin{figure}
\includegraphics[width=14.0cm]{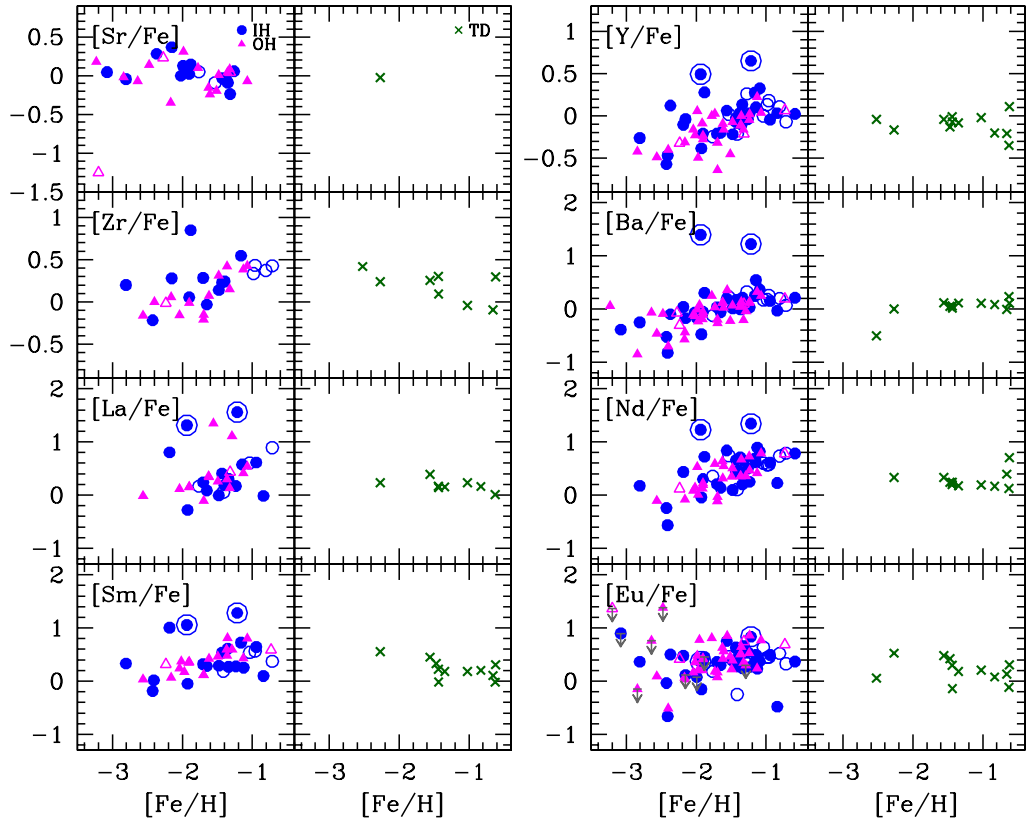}
\caption{Same as Figure \ref{fig:fepnc1_mem}, but for neutron-capture elements 
(Sr, Y, Zr, Ba, La, Nd, Sm, and Eu). The two sample stars, G~18--24 and 
BD$+$04\arcdeg 2466, which 
are identified as s-process rich stars, are marked with larger circles. }
\label{fig:fepnc2_mem}
\end{figure}

The [Sr/Fe] abundances are near-solar with scatter 0.06 to 0.24 dex
for both of the inner halo and the outer halo subsamples (the top panel 
of Figure \ref{fig:fepnc2_mem}). 
Although only one thick disk star has the Sr abundance measurement, 
the abundance of this star seems to agree with the inner/outer halo stars 
with similar metallicity. 
 The trends for the inner halo and the outer halo stars are indistinguishable. 
The near-solar [Sr/Fe] in [Fe/H]$>-2.0$ 
is in agreement with previous studies \citep{gratton94,burris00}, and the 
large dispersion in lower metallicity is also consistent with previous studies 
\citep{mcwilliam95,burris00,honda04,francois07}, 
although the sample size with [Fe/H]$<-2.0$ in this study is not large 
enough to quantify the scatter.

For [Y/Fe] ratios, 
the three subsamples show relatively large 
abundance scatter in all metallicities. We note that 
two stars, BD$+$04\arcdeg 2466 and G~18--24, with exceptionally 
high [Y/Fe], are the Y and Ba rich stars reported previously 
\citep{burris00,zhang09,ishigaki10} and marked in 
Figure \ref{fig:fepnc2_mem} with 
large open circles. The former was found to be a spectroscopic binary 
\citep{jorissen05}, while the binary nature of the latter  is unclear
 \citep{latham02}. 

The different behavior in the thick disk, inner halo and outer halo
 subsamples is modestly 
apparent in [Fe/H]$>-1.5$. The thick disk stars 
show the [Y/Fe] slightly below the solar value 
with a scatter of 0.14 dex. 
On the other hand, some of the inner halo stars show 
the [Y/Fe] above the solar value. Finally, 
the outer halo stars show, again, near to subsolar [Y/Fe]. 
In the lower metallicity, the inner halo and the outer halo 
subsamples both seem to show decreasing [Y/Fe] ratios 
with decreasing [Fe/H]
with a large scatter of $\ge 0.20$ dex.

Our results for the thick disk stars are generally consistent with 
the previous studies of \citet{bensby05}, \citet{prochaska00}, and 
\citet{reddy06}. 
The decreasing [Y/Fe] for the halo stars with decreasing [Fe/H]
is also consistent with 
the results of \citet{gratton94} and \citet{francois07}.

[Zr/Fe] ratios in the range [Fe/H]$>-1.5$ show a hint of 
difference among the subsamples 
in that the inner/outer halo stars seem to be more enhanced with 
[Zr/Fe] than the thick disk stars. In the lower metallicity, [Zr/Fe] is 
higher in the thick disk stars than in the halo stars.
The [Zr/Fe] trend with [Fe/H] for the halo stars is in agreement with 
that reported by \citet{gratton94}. 
Such a trend in [Zr/Fe] ratios toward lower [Fe/H]
has also been reported by 
\citet{francois07}. In this study, we show that the 
decreasing trend toward lower [Fe/H] is seen only for the inner/outer halo 
stars with [Fe/H] $< -2.0$, while it is not seen for 
the thick disk stars.

In the solar metallicity, a large fraction of Sr, Y, and Zr, 
is thought to be synthesized through s-process in low-to-intermediate 
mass AGB stars. In the lower metallicity, it has been suggested 
that another nucleosynthesis component is needed to explain observed abundance 
patterns of extremely metal-poor stars \citep[e.g.,][]{honda06}.   
We discuss this point by taking the ratio of these elements to Ba 
in Section \ref{sec:baratios}.

\subsubsection{Heavy neutron-capture elements: Ba, La, Nd, Sm, and Eu}

Similar to the behavior seen in the [Y/Fe]-[Fe/H] diagram, 
two of the sample stars, G 18--24 and BD$+$04\arcdeg 2466, stand out in 
abundance ratios for Ba, La, Nd, and Sm, which may suggest that 
surface composition has been modified during their 
internal evolution. 

All of the three subsamples show a similar [Ba/Fe] trend 
with [Fe/H]; [Ba/Fe] is near solar in the range [Fe/H]$>-1.5$, 
while it decreases toward lower [Fe/H]. 
The near-solar [Ba/Fe] ratios for the thick disk stars are in agreement
with previous studies \citep{prochaska00,reddy06}.
The observed decreasing [Ba/Fe] with decreasing [Fe/H] observed
for the halo stars is also consistent with the trend reported by 
\citet{gratton94} and \citet{francois07}.

For the [La/Fe], [Nd/Fe], and [Sm/Fe] ratios, a hint of a different 
trend with [Fe/H] between the thick disk and the inner/outer halo 
subsamples is modestly seen; the abundance ratios for the thick disk 
stars show a flat trend with [Fe/H] at values close to the 
solar ones, while the ratios for the inner/outer halo stars 
seem to increase with increasing [Fe/H]. 
In particular, at the highest metallicities, the 
inner/outer halo subsamples show clear overabundance 
at 0.45$\sim 0.60$ dex on average, which is not seen in 
the thick disk subsample  (see Table \ref{tab:mean_dev}).
In the study of \citet{mashonkina04}, who performed Nd abundance 
analysis of 60 thin disk, thick disk and halo stars, using a 
spectral synthesis techniques, such an abundance difference between 
the thick disk and halo stars was not identified. Instead, 
they reported that a similar overabundance of [Nd/Fe] in the range 
0.23 to 0.45 dex for both thick disk and halo stars exists in their overlapping 
metallicity range. We note that some offset in abundances between the 
two studies exists since \citet{mashonkina04} employed empirical $\log gf$ 
values while this study adopts the $\log gf$ values from \citet{ivans06}, 
which were originally taken from the measurement by \citet{denhartog03}. 
Further precise studies on the heavy neutron-capture elemental abundances 
between the thick disk and halo stars in their overlapping 
metallicity range with a larger sample are desirable 
to convincingly conclude on the abundance differences among the different 
Galactic populations.  

\subsubsection{[X/Ba]-[Ba/H] diagrams for neutron-capture elements}
\label{sec:baratios}

\begin{figure}
\includegraphics[height=11.0cm]{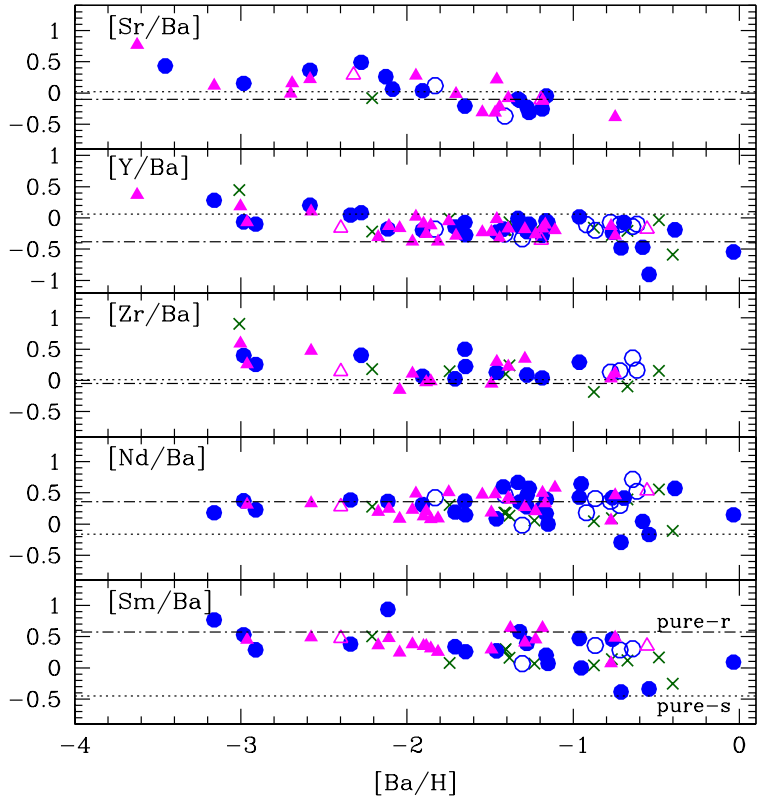}
\caption{[Sr/Ba], [Y/Ba], [Zr/Ba], [Nd/Ba] and [Sm/Ba] ratios plotted 
against [Ba/H]. The symbols are the 
same as in Figure \ref{fig:fepnc1_mem}. The abundance ratios of the solar system s-process (dotted) and r-process (dash-dotted) components predicted by \citet{arlandini99} are indicated by horizontal lines.}
\label{fig:sr_y_zr_nd_sm_ba_mem}
\end{figure}

Abundance ratios among neutron-capture elements are 
widely used as a diagnostic for possible nucleosynthesis 
sites of these elements \citep[e.g.][]{francois07}.
Figure \ref{fig:sr_y_zr_nd_sm_ba_mem} shows [Sr/Ba], 
[Y/Ba], [Zr/Ba], [Nd/Ba], and [Sm/Ba] ratios plotted against [Ba/H] 
for the thick disk, inner halo and outer halo subsamples. 
The values for the solar system s- and r-process components from 
\citet{arlandini99} are indicated 
by the dotted and dash-dotted lines, respectively.

In the range [Ba/H]$<-2.0$, abundance ratios of 
the lighter neutron
capture elements, Sr, Y, and Zr, for
most of the sample stars show higher 
values than those for the solar-system r-process component 
and an increasing trend toward lower [Ba/H]. 
Both the inner halo and the outer halo stars, as well as 
thick disk stars seem to follow the above trend. 
Such trend has previously been reported by 
\citet{francois07}, in which it was interpreted as 
a possible indication that an extra production mechanism for lighter 
neutron-capture elements over heavier ones is required. 
The mechanism likely responsible for the excess of 
light neutron capture elements 
is frequently called the Light Element Primary Process (LEPP).
Their nature and astrophysical sites are 
investigated by nucleosynthesis calculations in core-collapse SNe 
\citep{wanajo11,arcones11}.   
Our results imply that the mechanisms responsible for the Sr, 
Y, and Zr enrichment were commonly efficient 
among the progenitor of the thick disk, inner halo and outer halo 
stars with [Ba/H]$<-2.0$.

In [Ba/H] above $\sim -1.5$, [Nd/Ba] and [Sm/Ba] 
ratios for a majority of the thick disk stars and 
some inner/outer halo stars are below the 
solar-system r-process values. This result is consistent with the 
previous study of \citet{mashonkina04}. 
A possible indication is that the s-process may have contributed to the 
chemical enrichment in progenitors of both  
the thick disk and inner/outer halo components.  
In the next subsection, we will discuss the r- and s-process 
contribution on the observed
abundances of neutron-capture elements in our sample based on the 
Eu abundances.

\subsubsection{Europium}
\label{sec:eu}

The bottom-right panels of Figure \ref{fig:fepnc2_mem} show the [Eu/Fe] 
abundance ratios plotted against [Fe/H] for the thick disk, 
inner halo and outer halo subsamples. 
The most remarkable feature in these plots is that 
the [Eu/Fe] ratios are more enhanced in the inner/outer halo 
subsamples than in the thick disk subsample by $\sim0.2-0.3$ dex 
in the metallicity range of [Fe/H]$>-1.5$.
In Table \ref{tab:mean_dev}, the mean values for the three subsamples 
taking into account only dwarfs ($\mu_{\rm d}$) or giants ($\mu_{\rm g}$) are also indicated.  
The $\mu_{\rm d}$ for [Eu/Fe] in the inner and outer halo subsamples 
is again larger than that in  the thick disk subsample at metallicities 
[Fe/H]$>-1.5$, while differences in $\mu_{\rm g}$ among the three 
subsamples are not clear.
Although further confirmation with a larger sample is desirable  
for a definite conclusion, the difference in $\mu_{\rm d}$ 
may indicate that the abundance difference is not totally caused by systematic 
errors due to the inclusion of both dwarfs and giant stars in our sample.

\begin{figure}
\includegraphics[height=11.0cm]{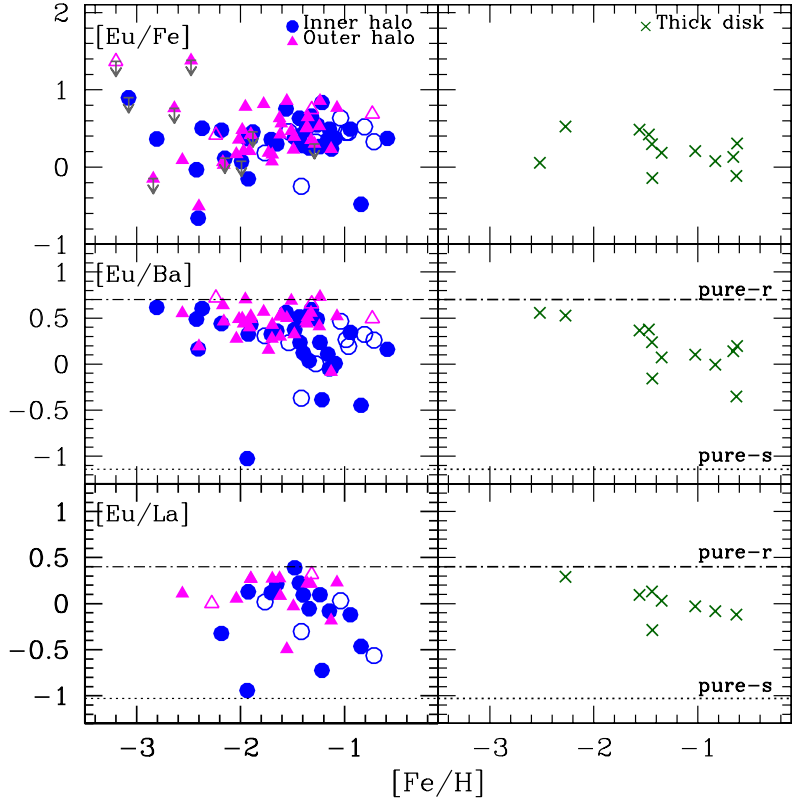}
\caption{Same as Fig \ref{fig:fepnc1_mem} but for [Eu/Fe], [Eu/Ba], 
and [Eu/La]. The abundance ratios of the solar system s-process 
(dotted) and r-process (dash-dotted) components predicted by 
\citet{arlandini99} are indicated by horizontal lines.}
\label{fig:eu}
\end{figure}

To examine whether the enhanced [Eu/Fe] ratios 
observed in the inner and outer halo 
stars are caused by an excess of the r- or s-process 
contribution to the production of Eu, the middle and the 
bottom panels of Figure \ref{fig:eu} show 
the [Eu/Ba] and [Eu/La] ratios in the sample stars
 as well as the values for the pure r- and s- process components in 
the solar system material predicted by \citet{arlandini99}.  
The [Eu/Ba] ratios in many of the inner halo and outer halo subsamples
show a flat trend with [Fe/H] up to [Fe/H]$\sim -1.5$ and do not 
significantly deviate from the value for the solar-system r-process 
component ([Eu/Ba]$\sim0.7$).  
In the same metallicity range, the
thick disk subsample, on the other hand, predominantly  
shows the [Eu/Ba] ratios below the flat inner/outer halo trend. 

Similar behavior can be seen in the [Eu/La] versus [Fe/H] plot. 
Again, in the metallicity range of [Fe/H]$>-1.5$, 
the thick disk stars tend to show lower [Eu/La] ratios than the typical 
inner/outer halo stars.  Some of the inner/outer halo 
stars including the known s-rich stars show much lower [Eu/Ba] and 
[Eu/La] ratios 
close to the values for the solar-system s-process component. 

These results suggest that Eu in our inner/outer halo 
sample stars is predominantly an r-process origin. 
In contrast, the observed lower [Eu/Ba] and [Eu/La] ratios 
in the thick disk stars may  
indicate contribution of the s-process to the chemical 
enrichment during the formation of the thick disk. 
A similar conclusion has been reached for the thick disk stars 
by previous studies 
\citep{mashonkina00,bensby05}.

\begin{figure}
\includegraphics[height=11.0cm]{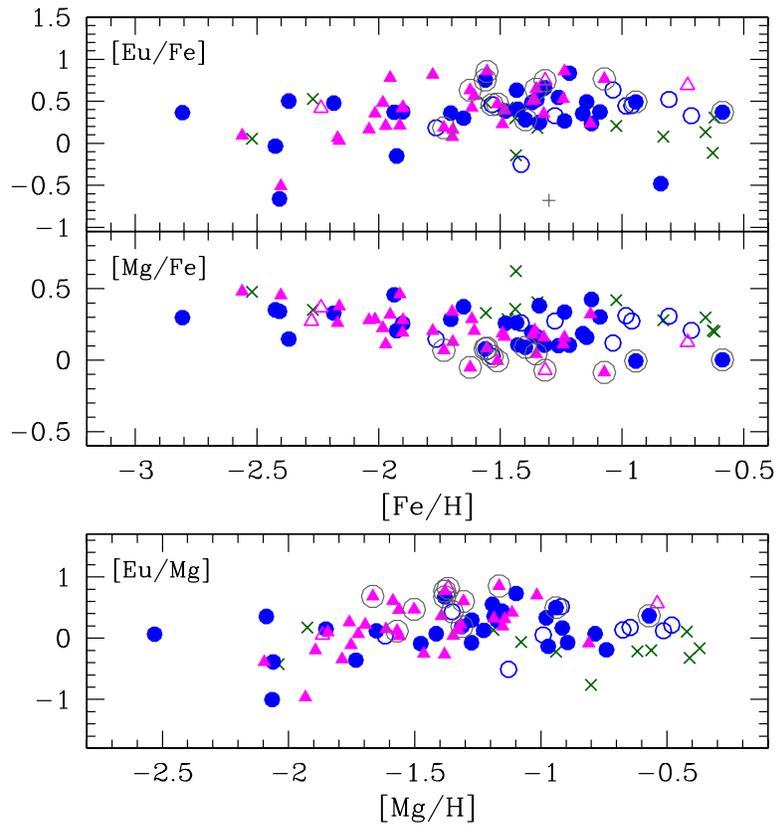}
\caption{[Eu/Mg] ratios plotted against [Mg/H]. The symbols are the 
same as in Fig \ref{fig:fepnc1_mem}.}
\label{fig:eu_mg}
\end{figure}

If Eu is an r-process origin and if the r-process elements are
 predominantly ejected in Type II SNe as $\alpha$ elements, 
it is puzzling that the trend found in the [Eu/Fe] 
ratios for these stars does not follow that seen in the $\alpha$ elements.
As can be seen in the top and middle panels of Figure \ref{fig:eu_mg},
while the [Mg/Fe] ratios in many of the inner/outer halo stars 
follow a decreasing trend with increasing [Fe/H] (Paper I),  
the [Eu/Fe] ratios for these stars do not. This result is in contrast 
to the expectation that the [Mg/Fe] and [Eu/Fe] ratios would show a similar 
trend with [Fe/H] if Mg and Eu are produced in the same astrophysical site, 
presumably Type II SNe of massive stars. 
Rather, the stars with low-[Mg/Fe] ratios (gray circled symbols) in 
our sample tend to show higher [Eu/Fe] ratios. 

The difference in the behaviors of the Eu and Mg abundances 
can be more clearly seen in the bottom panel of Figure \ref{fig:eu_mg}, 
where the [Eu/Mg] ratios are plotted against 
[Mg/H]. In this plot, the thick disk stars show 
almost a flat [Eu/Mg] trend for all of the [Mg/H] range at about the 
solar value, while the inner/outer halo stars exhibit 
larger values of [Eu/Mg], mildly increasing with increasing [Mg/H]. 
The similar [Eu/Mg] enhancement has been reported previously by 
\citet{letarte10} for stars in Fornax dwarf spheroidal galaxy. 
Unlike the present sample, the Fornax stars were found to be rich in 
s-process elements like Ba, and thus the enhanced [Eu/Mg] ratios 
have been interpreted as due to the excess of the s-process in this galaxy 
\citep{letarte10}. A different interpretation is likely needed
to explain the high [Eu/Mg] for the present sample since 
the signature of significant s-process enrichment like that seen in 
Fornax is not observed in these stars as mentioned above.  

The implication of these results are twofold; (1) on the astrophysical 
site of the r-process and (2) on the different chemical enrichment 
histories in the progenitors of the thick disk and the inner/outer 
stellar halos. 
For point (1), the results favor the scenario that the astrophysical 
site which produces Eu is different from that for Mg. 
Magnesium is thought to be largely synthesized in 
massive stars during the hydrostatic burning and 
ejected through Type II SNe, while 
more than 90 \% of Eu in the solar system material is thought to be 
synthesized in the r-process \citep{arlandini99}, whose astrophysical 
site is still unknown.  
A Type II SN of a relatively low-mass progenitor 
(8-10 $M_{\odot}$) is suggested to be the primary site for 
 the r-process \citep[e.g.][]{wheeler98}. 
These low-mass SNe as well as higher mass $>20 M_{\odot}$ SNe are
considered in the Galactic chemical evolution model and are shown 
to be consistent with observations of metal-poor stars \citep{ishimaru99}. 
Alternatively, neutron-star mergers are 
also suggested as a possible r-process site \citep{freiburghaus99}. 
Each of these scenarios suffer from large theoretical uncertainties 
and thus the r-process site remains controversial 
\citep[e.g.][]{argast04}.
The observed signature of the different trends in the [Eu/Fe] and the [Mg/Fe] 
ratios appear to be consistent with the scenario that 
Type II SNe of different progenitor mass ranges are responsible 
for the production of dominant Eu ($8-10 M_{\odot}$) and 
Mg (e.g. $>10 M_{\odot}$), which was also suggested from chemical evolution 
models \citep[e.g.][]{tsujimoto98}. However, the
 high [Eu/Mg] ratios at [Fe/H]$> -1.5$ 
seen in the inner/outer halo stars 
may require that the Eu production have a much longer timescale 
than Mg. Whether an alternative scenario such as neutron-star mergers 
can explain the abundance results must be tested by chemical 
evolution models taking into account the star formation history 
in possible progenitor systems of 
the thick disk and stellar halo.

For point (2), on the chemical enrichment histories 
in the progenitors of the thick disk and stellar halo, 
Paper I showed that the [Mg/Fe] ratios in the inner/outer halo stars 
show a decreasing trend with [Fe/H] and lower than in the thick disk stars. 
Low [Mg/Fe] ratios in stars with [Fe/H]$>-1.5$ are frequently 
interpreted as the result of a larger contribution of Fe from Type Ia SNe 
in these stars, presumably due to the lower star formation rate 
in their progenitor systems. However, 
rather enhanced [Eu/Fe] ratios found in the low [Mg/Fe] stars in our sample
 may indicate difficulty in interpreting their abundances
 by this simple picture alone. This is because the larger contribution of Fe 
would also reduce the [Eu/Fe] ratios in these low [Mg/Fe] stars 
than in the thick disk stars, which is in contrast to the observed 
trend. Chemical evolution models that allow variation of, not only star formation 
rates, but also other factors such as an IMF may be 
necessary to explain the observed behavior of [Eu/Fe] and [Mg/Fe] ratios
consistently among the thick disk and halo stars.

\subsection{Abundance kinematics correlation}
\label{sec:abu_kin}

\subsubsection{Abundance patterns of stars with extreme kinematics}
\begin{figure}
\includegraphics[width=14cm]{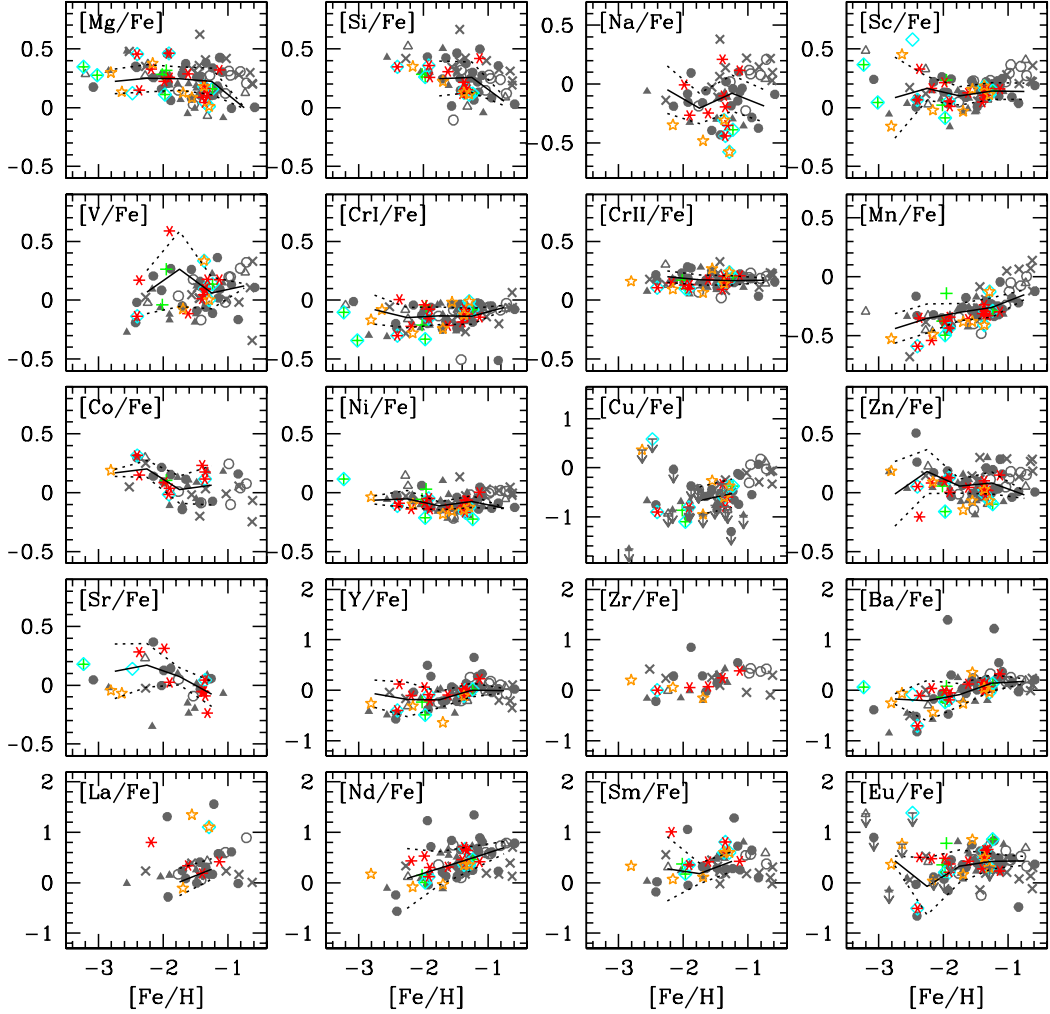}
\caption{[X/Fe]-[Fe/H] diagram for stars with extreme kinematics: 
 stars with extreme rotational 
velocities ($V_{\phi}<-150$ or $>250$ km $s^{-1}$; stars), high maximum distance 
from the Galactic plain ($Z_{\rm max}>20$ kpc; crosses),  
large apocentric distance ($R_{\rm apo}>30$ kpc; diamonds) 
and orbital eccentricity close to unity 
($e>0.95$; astarisks). The other sample stars are shown 
in gray symbols. The solid line in each panel connects 
mean abundance of the inner halo stars with $P_{\rm IH}>0.95$ 
in each of 0.5 dex [Fe/H] intervals. The dotted lines
indicate the range within 1$\sigma$ scatter about the means. }
\label{fig:abundance_extremekin}
\end{figure}

Figure \ref{fig:abundance_extremekin} highlight the abundances of 
stars with extreme 
orbital parameters: i.e., stars having extreme rotational 
velocities ($V\phi<-150$ or $>250$ km s$^{-1}$: stars), high maximum distance 
from the Galactic plane ($Z_{\rm max}>20$ kpc: crosses),  
large apocentric distance ($R_{\rm apo}>30$ kpc: diamonds), 
and eccentricity close to unity 
($e>0.95$: asterisks). 
Stars that meet more than one of the four categories are shown 
with overlapping symbols. Mean abundances of
 the typical inner halo stars with $P_{\rm IH}>0.95$ 
in each metallicity bin of 0.5 dex 
are shown in the solid lines, where the dotted lines show 
a range in mean$\pm$ dispersion if 
there are $\ge 2$ stars in each [Fe/H] interval.  The  
extreme cut of $P_{\rm IH}>0.95$ was employed to exclude 
possible contamination of stars with thick disk-like or 
outer-halo-like kinematics. 

It can be seen that in the metallicity below [Fe/H]$\sim -2.0$, 
stars with any of the extreme orbital parameters well 
overlap with the typical inner halo distribution.
On the other hand, at higher metallicities ([Fe/H]$\gtrsim -2$), 
deviation from the 
typical inner halo distribution for some elements is identified. 
Both stars with extreme rotational velocities and 
large apocentric distance show 
signatures of lower [Mg/Fe], [Na/Fe], [Ni/Fe], and 
[Zn/Fe] ratios than the other stars 
in [Fe/H]$>-2.0$. 
The stars with orbital eccentricity close to 
unity seems to be more similar to the normal halo 
stars.

These results imply a possible correlation of 
chemical abundances with some of the orbital 
parameters in [Fe/H]$\gtrsim -2.0$, while no correlation 
is expected in the lower [Fe/H]. The abundance-kinematics 
correlation in the metallicity range $-1.5<$[Fe/H]$<-0.5$ is examined in the 
next subsection.

\subsubsection{Abundance ratios vs. orbital parameters}
In the previous subsections, we show the differences in elemental 
abundances in the range [Fe/H]$>-2.0$ between stars with extreme kinematics 
and the stars having typical inner halo kinematics. 
In order to examine whether the observed differences came from 
any correlation between [X/Fe] with the orbital parameters 
among the inner and outer halo stars, 
we calculate the linear correlation coefficient for the [X/Fe] 
with $V_{\phi}$, $Z_{\rm max}$, $R_{\rm apo}$, and orbital eccentricity $e$ 
and probability 
at which a null hypothesis of no correlation can be rejected. 
Some of the [X/Fe] ratios appear to be weekly correlated with 
[Fe/H]. In order to reduce the effect of the [X/Fe] versus [Fe/H] 
correlation in examining the [X/Fe] versus orbital parameter correlation,   
we limit the sample stars to those with metallicity $-1.5<$[Fe/H]$<-0.5$. 
Elements 
for which more than 20 stars are available in the above metallicity 
range are considered. 
As a result, we found that for Na, Sc, Ni, and Zn, the possible 
correlation is suggested for one or more of the above orbital 
parameters with the probability for the 
null hypotheses of less than 
5\%. 

\begin{figure}
\includegraphics[width=14.0cm]{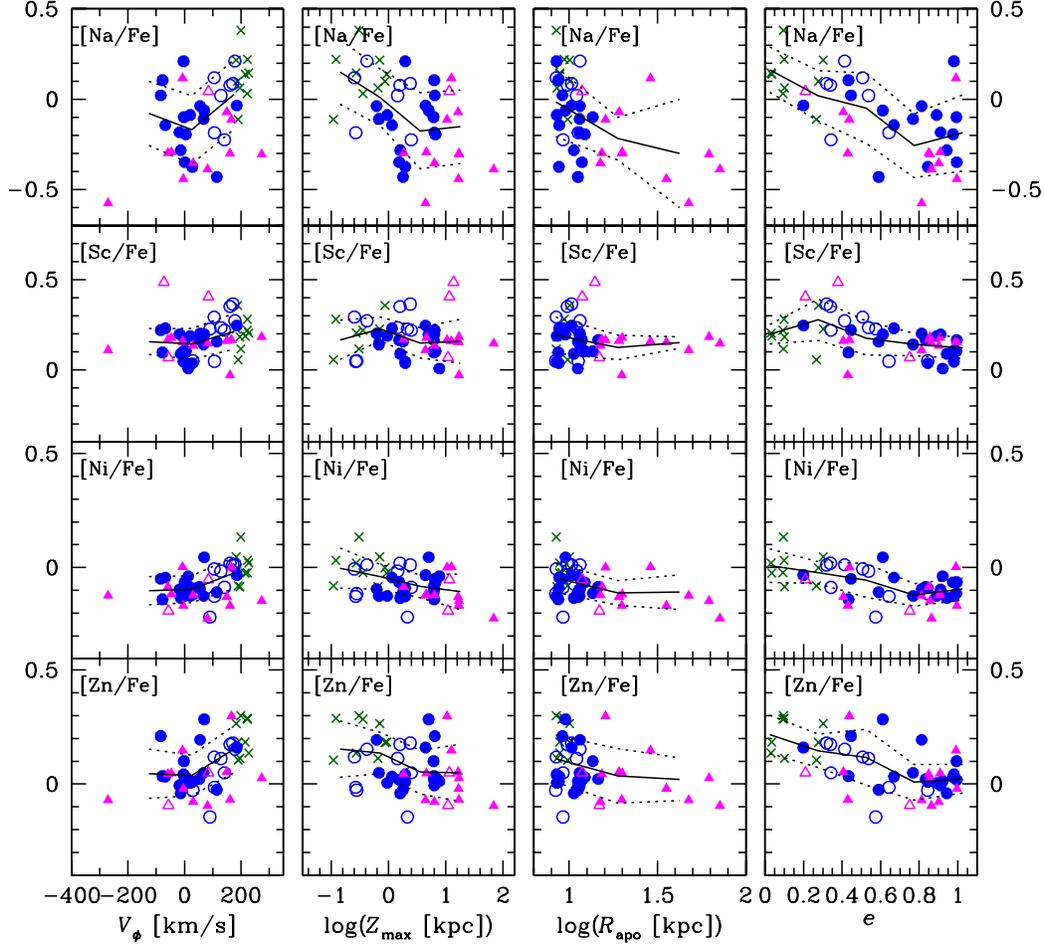}
\caption{From top to bottom: [Na/Fe], [Sc/Fe], [Ni/Fe], and [Zn/Fe] plotted 
against orbital parameters. From left to right: $V_{\phi}$, $\log Z_{\rm max}$, $\log R_{\rm apo}$, 
and orbital eccentricity in the metallicity range $-1.5<$[Fe/H]$\leq -0.5$. 
The mean values and their 1$\sigma$ scatter in each bin of the orbital 
parameters are indicated with solid and dotted lines, respectively.     
The symbols are the 
same as in Fig \ref{fig:fepnc1_mem}.
}
\label{fig:na_sc_ni_zn_kin}
\end{figure}

Figure \ref{fig:na_sc_ni_zn_kin} plots (top to bottom) 
[Na/Fe], [Sc/Fe], [Ni/Fe] and [Zn/Fe] 
as a function of (left to right) $V_{\phi}$, $\log Z_{\rm max}$, 
$\log R_{\rm apo}$, 
and $e$ for the sample stars with 
$-1.5<$[Fe/H]$\le -0.5$.  
As in Figure \ref{fig:fepnc1_mem}, 
the crosses, filled circles, and filled triangles indicate
the thick disk stars, the inner halo stars, and 
the outer halo stars, respectively. Their 
intermediate populations are shown in the open symbols 
(thick disk/inner halo: open circles, inner halo/outer halo
: open triangles).

In the [X/Fe] versus $V_{\phi}$ plot, the stars on retro-grade 
orbit ($V_{\phi}<0.0$) tend to show lower abundance ratios compared to the 
stars with prograde orbit including thick disk stars (cross) and thick disk/
inner halo intermediate stars (open circles). One outer halo star
with an extreme prograde rotation ($V_{\phi}\sim 280$ km s$^{-1}$) 
clearly shows a different abundance pattern compared to the 
thick disk stars. 

In the [X/Fe] versus $\log Z_{\rm max}$ plot, the abundance scatter 
appears to be larger in $Z_{\rm max}>1$ kpc. The $Z_{\rm max}\sim 1$ kpc 
corresponds to the transition between the thick disk dominant to 
the inner halo dominant region according to \citet{carollo10}. At larger 
$Z_{\rm max}$, the trend is not very clear for all of the elemental 
abundances.

In the [X/Fe] versus $\log R_{\rm apo}$ plot, stars with $R_{\rm apo}>30$ 
kpc clearly show lower [Na/Fe] and [Ni/Fe] ratios compared to the 
other halo stars with smaller $R_{\rm apo}$. 

Interestingly, the correlation between the [X/Fe] and $e$ can be 
seen for all of the four elements. The calculated probability 
for the null hypothesis of no correlation is less than 2 \% for 
these elements. 
As mentioned in the previous subsection, the highest $e$ stars 
consist of both the inner and outer halo stars in our sample. 
This might partly reflect our selection of high-velocity stars 
in the solar neighborhood, which preferentially select high-eccentricity 
stars. The tendency for these stars to have different chemical 
properties may result from our selection bias toward 
stars accreted from other systems, although such population may 
not be dominant in the inner halo region as suggested by 
\citet{carollo10} 

\section{Summary and discussion}
\label{sec:discussion}

We estimate elemental abundances of sodium, iron-peak, 
and neutron-capture elements of
97 dwarf and giant stars kinematically belonging to the MW thick disk, 
inner halo, and outer halo components 
in the metallicity range of $-3.3<$[Fe/H]$<-0.5$.
Using these sample, characteristic trends in 
the [X/Fe]-[Fe/H] diagrams for each of the 
three subsamples are
investigated as a clue to the formation mechanisms of 
these old Galactic components.
Our results show that the abundances are largely similar among the thick disk, 
inner halo, and outer halo subsamples in the metallicity range of [Fe/H]$<-2$. 
In contrast, the abundance differences for some elements among the 
three subsamples are identified at higher metallicities ([Fe/H]$\gtrsim-1.5$).
Our main results are summarized as follows.

\begin{itemize}
\item The inner halo and outer halo stars show lower 
[Na/Fe], [Ni/Fe], [Cu/Fe], and [Zn/Fe] ratios than the thick disk stars 
in the metallicity range of [Fe/H]$>-1.5$. 
In particular, the sample stars with relatively low [Mg/Fe] show low 
[Zn/Fe] ratios compared to the other halo stars, which is in line with 
the results of \citet{nissen11}. 
\item All of the three subsamples show an increasing [Mn/Fe] trend with 
[Fe/H], which may suggest that Mn from Type Ia SNe has contributed 
to enriching the progenitors of the thick disk, inner and outer stellar halos.   
\item The [Eu/Fe] ratios in the inner/outer halo stars are 
higher than in the thick disk stars in the range [Fe/H]$>-1.5$. 
This behavior is in contrast to that seen in
[Mg/Fe] ratios, for which many of the inner/outer halo stars show lower 
values than the thick disk stars. 
The different behavior of [Eu/Fe] and 
[Mg/Fe] ratios for the three subsamples may 
imply that the production site of Eu is different from that of Mg, although 
more exact interpretations require detailed chemical evolution 
modelling in possible progenitors of the thick disk and stellar halos 
as well as determination of yields for these elements.
\item The [Eu/Ba] and [Eu/La] ratios for the thick disk stars and 
a handful of the inner halo stars are below 
the values expected for the solar-system r-process component 
\citep{arlandini99}, which may indicate contribution of the s-process 
from low-to-intermediate mass AGB stars to the
chemical enrichment in the progenitors of these stars.
On the other hand, many of the inner and outer halo 
stars follow the flat trend up to [Fe/H]$\sim -1.5$, likely 
suggesting that the r-process dominates in synthesizing heavy 
neutron-capture elements in their progenitor systems.  
\item The inner/outer halo stars with extreme retrograde rotation, 
large $Z_{\rm max}$, and large $R_{\rm apo}$ tend to show 
lower [Na/Fe], [Ni/Fe], and [Zn/Fe] ratios 
than those with normal inner halo kinematics in [Fe/H]$>-2$. 
In the lower metallicities the abundances are almost indistinguishable 
among the kinematically different populations.  
\item In our inner and outer halo subsamples with 
metallicities $-1.5<$[Fe/H]$\le -0.5$, signatures of the correlation 
between [Na/Fe], [Sc/Fe], [Ni/Fe], and [Zn/Fe] ratios and the orbital 
eccentricity are identified.
\end{itemize}

Based on the above results,  we discuss the possible formation 
mechanisms of the thick disk, inner and outer stellar halos, with particular 
focus on what are the likely progenitors of these components.

\subsection{The thick disk}

Several scenarios from theoretical models for 
the thick disk formation are proposed; 
(1) monolithic dissipative collapse 
of a disk component \citep[e.g.][]{burkert92}, (2) the mergers of 
satellites that are tidally disrupted to populate the 
thick disk \citep[e.g.][]{abadi03}, (3) heating a pre-existing 
thin disk through numerous mergers of dark matter subhalos or dwarf galaxies 
\citep[e.g.][]{quinn93,velazquez99,hayashi06}, (4) multiple 
dissipative mergers of building blocks that trigger 
rapid star formation \citep[e.g.][]{brook04,brook05}, 
and (5) early evolution of a gas-rich clumpy young disk 
\citep[e.g.][]{bournaud09}. 
It has also been proposed that (6) the secular radial migration of the 
thin disk may be responsible for the thick disk formation 
\citep[e.g.][]{haywood08,schonrich09}. 
Which of the above mechanisms is dominant and 
how the formation of the thick disk relates to the formation 
of the other Galactic components, namely, the bulge, thin disk, 
and stellar halo, are still controversial.  

In the present study and in paper I., we found the 
remarkable abundance difference for several elements 
between the thick disk and the inner/outer halo stars 
in their overlapping metallicity range 
($-1.5\lesssim$[Fe/H]$\lesssim -0.5$). 
The main implications from this result on the formation 
mechanisms of the thick disk component are summarized 
as follows; 
first, the high [Na/Fe] ratios, as well as 
high [Mg/Fe] ratios reported in Paper I., compared to 
many of the inner/outer halo stars indicate that  
chemical evolution of the thick disk was driven primarily 
through nucleosynthesis products from Type II SNe. 
 This may suggest that the thick disk formation timescale was 
sufficiently short and/or the IMF in the progenitor of the thick disk 
was biased toward high mass 
stars so that nucleosynthesis products of low-mass stars (i. e., Type Ia 
SNe) played a minor role.  
Second, the increase in [Mn/Fe] with [Fe/H] for the thick disk 
stars indicates that some contribution of Type Ia SNe may present, 
since Mn is predominantly synthesized in the explosive burning 
of Type Ia SNe in the metallicity typical of the thick disk stars 
\citep{kobayashi11}. 
Third, the [Eu/Ba], [Eu/La], [Sm/Ba], and [Nd/Ba] ratios for the thick disk 
are below the values expected for the solar system r-process components, 
which suggests that the s-process has contributed to the 
chemical enrichment of the thick disk. 
Fourth, the lower [Eu/Fe] and [Eu/Mg] ratios in the thick disk 
stars than in the inner/outer halo stars further support
distinct chemical enrichment histories among these components, 
while its interpretation depends on a currently unknown 
astrophysical site for Eu production.
Finally, for most of the 
elements, scatter in the abundance ratios 
for the thick disk stars is comparable to or smaller than the observational 
errors, which is in contrast to that for the inner/outer halo stars with 
[Fe/H]$>-1.5$. 
The lack of scatter may indicate
 that the gas from which the thick disk has formed 
may have been relatively well mixed.

The short timescale for the formation of the thick disk 
might be expected in scenarios (1), (4), and (5), where 
sufficient cold gas is supplied 
in the form of smooth gas accretion \citep{bournaud09} 
or multiple gaseous mergers \citep{brook04} 
at high redshift.   
Scenario (3) and (6) could also be possible if a 
preexisting thin disk formed under the sufficiently 
high star formation rate so that the whole disk stars rapidly 
became metallicity as high as the thick disk stars. 
In each of the above cases, the nucleosynthesis products of 
Type Ia SNe and s-process should have been rapidly  mixed in 
the ISM. 
Consistent with previous studies, the abundances of the 
thick disk stars in our sample do not resemble those of the known 
dwarf satellite galaxies around the MW, 
like Fornax, whose abundance is characterised by much lower 
[$\alpha$/Fe] and [Na/Fe] and higher [Ba/Fe] at similar [Fe/H]
 \citep{letarte10}. Therefore, it is unlikely that the 
thick disk has been totally built through an assembly of dSphs having 
similar properties as surviving Galactic dSphs (e.g. scenario (2)).

The four thick disk stars with [Fe/H]$<-0.8$ in our sample, all of which 
have kinematic properties similar to the proposed MWTD component, have 
particular implication on the earliest evolution of the MW thick disk. 
As shown in the previous sections and in Paper I., 
the candidate MWTD stars show 
distinct abundance ratios for several elements 
from those of the inner/outer halo stars at metallicities [Fe/H]$\gtrsim -1.5$. 
Even though the abundance ratios in the candidate 
MWTD become almost indistinguishable
from the inner/outer halos at lower metallicities, 
their trends with [Fe/H] seem to smoothly follow those seen in the 
canonical thick disk stars 
(i.e. the thick disk stars with $-0.8<$[Fe/H]$<-0.5$) 
with relatively small scatter. This result implies that 
these candidate MWTD stars may have indeed formed within the progenitors of 
the thick disk, which are distinct from those of the stellar halo. 
In other words, our result favors the interpretation that 
they are remnants of the ancient MW disk system 
rather than the interloper of the halo population that
happened to acquire disk-like kinematics. A possible implication 
is that, whatever the formation mechanism is, the progenitor of the 
thick disk may have been a member of a metal-poor disk system, 
which later 
experienced relatively rapid and homogeneous chemical evolution. 
Further studies on the detailed chemical abundances versus kinematics 
as well as ages of individual metal-poor stars are necessary to get deeper insight 
into the formation of the thick disk and the origin of its MWTD component.

\subsection{The stellar halos}

One of the key questions on the global formation and evolution 
of the MW stellar halos is whether the halos have been assembled from 
smaller stellar systems similar to the dSph galaxies currently orbiting 
the MW. In the following subsections, we first address this question by 
examining whether the observed abundances in our inner/outer halo 
subsamples resemble those reported for stars in classical as well as 
ultra-faint dwarf galaxies. For this comparison, we separately 
consider two metallicity ranges ([Fe/H]$>-2$ and $<-2$) since 
observed abundance trends for
 the inner/outer halo subsamples vary depending on 
their metallicity.  
Then, we discuss whether these abundances in the MW thick disk, 
inner- and outer stellar halos are compatible with 
proposed MW formation scenarios based on
 other observations or numerical simulations.

\subsubsection{Comparisons with MW dSph galaxies}

{\it [Fe/H]$<-2$}.  In metallicities below $-2.0$, the abundances of 
several elements in both the inner and outer halo 
subsamples are found to be in general agreement with known classical 
and ultra-faint dwarf galaxies regardless of their kinematics. 
In this metallicity range, 
the inner/outer halo stars generally show enhanced [Mg/Fe] and [Si/Fe] ratios, 
near-solar ratios for [Ni/Fe] or [Sc/Fe], and the low [Mn/Fe] ratios. 
Similar abundance ratios
 were reported for the lowest metallicity stars ([Fe/H]$\lesssim-2.5$) 
in classical dwarf galaxies 
like Draco \citep{cohen09},  
Ursa minor \citep{cohen10}, Carina \citep{venn12}, and Sculptor \citep{starkenburg13}.
The abundances of light elements ($Z<30$) in 
newly discovered ultra-faint dwarf galaxies also show 
abundances similar to as the halo stars \citep{frebel10,simon10, gilmore13} in the 
whole metallicity range spanned by these stars ([Fe/H]$\lesssim -2.0$).

Despite these similarities, our sample of both inner and outer halo 
stars with [Fe/H]$<-2.0$ also confirms some of the known descrepancies
in abundances between the dSphs and MW halo stars. 
First, as discussed in Paper I., our sample of inner and outer halo 
stars both shows enhanced [Mg/Fe] ratios, while 
those in extremely metal-poor stars in Sextans dSph 
are systematically lower than the MW halo stars at similar 
metallicities \citep{aoki09} 
Second,  abundance ratios of neutron-capture elements like [Sr/Fe] or 
[Ba/Fe] tend to be low in both classical \citep{aoki09,venn12} and 
ultra-faint dwarf galaxies \citep{koch08,frebel10,simon10}, with an exception 
of a possible binary star \citep{honda11}, while the  
field halo stars including our inner/outer halo samples 
exhibit both high and low abundances of these neutron-capture elements. 
In conclusion, the abundance differences as indicated above remain
even if the present sample of the outer halo stars 
with extreme kinematics is taken into account. 
The suggested abundance differences in neutron capture elements 
may point to some differences in 
chemical evolution between progenitors of the MW
field halo and some of the dSphs. In order to get 
deeper insights into this discrepancy, chemical evolution 
modeling which takes into account star formation and mixing 
of elements within a small system as well as investigation of 
nucleosynthesis yields of neutron-capture elements at this low metallicity 
would be necessary.

To summarize, in the metallicity below [Fe/H]$\sim -2.0$, 
abundances of 
$\alpha$ and Fe peak elements in both of our inner and outer halo 
subsamples largely overlap with those of the extremely 
metal-poor stars in some of the classical and 
ultra-faint dwarf galaxies.
The similarity supports the hypothesis that metal-poor 
stars in both of the inner and outer halos in this 
low-metallicity range are at 
least in part accreted from systems similar to many of 
the currently known
classical and ultra-faint dwarf galaxies. 
We note that, as we have mentioned in Section \ref{sec:metallicity}, 
recent surveys suggest that metallicity distribution function 
is different between the inner and outer halos, and thus, 
the inner halo stars with [Fe/H]$\sim -2.0$ are 
relatively minor population. Therefore, the accretions of such
metal-poor dSphs-like systems are likely more dominant in the 
outer halo than in the inner halo.

{\it [Fe/H]$>-2$}. In this metallicity range, abundances of several 
elements in most of the 
outer halo stars, especially those having extreme kinematics, overlap 
with those in the MW dSph stars, while the stars with 
the typical inner halo kinematics typically show a larger discrepancy in 
the abundances from the dSphs as recognized in previous studies 
\citep{venn04,tolstoy09}. 
As an example, [Na/Fe] ratios in stars in Fornax and Sagittarius 
dSphs with $-1.5<$[Fe/H]$<-0.5$ have been reported to be subsolar in a range 
$-1<$[Na/Fe]$<-0.2$ \citep{letarte10,sbordone07}, while the present 
sample of the outer halo stars also show subsolar [Na/Fe] ratios in 
contrast to the thick disk and some of the inner halo 
stars mostly showing the supersolar ratios.
A similar trend is found in the [Ni/Fe] ratios, for which the outer 
halo stars largely show subsolar values similar to stars in 
Fornax and Sagittarius dSphs \citep{letarte10,sbordone07}, while 
the thick disk and some of the inner halo stars generally show 
supersolar values.

On the other hand, differences in Ba abundances 
in these dSph stars \citep{sbordone07,letarte10} 
and the inner/outer halo stars seem to be substantial 
regardless of kinematics. 
Ba in moderately metal-poor stars is thought to be mainly produced in 
low-to-intermediate mass AGB stars. The difference between the 
MW inner/outer halo stars and the dSphs indicates that chemical 
enrichment by low-to-intermediate mass stars may be different in 
the progenitors of the inner/outer halo from the 
surviving dSphs. 

Then, the intriguing question is, what is the origin of the kinematically 
classified outer halo stars with [Fe/H]$>-2.0$ which show 
chemical abundance ratios different from the normal inner halo stars? 
Again, we note that recent determination of metallicity 
distribution function \citep[e.g.][]{an13} suggests that the 
outer halo population 
is dominated by more metal poor stars ([Fe/H]$\sim-2.0$), and thus 
our kinematically defined outer halo stars with [Fe/H]$>-2.0$ 
could be a relatively minor population (see Section \ref{sec:otherobs}).
The implication from the abundance results in the present study and in Paper I 
is that the progenitors of these relatively metal-rich 
outer halo stars may have experienced 
a certain interval of chemical evolution that has allowed the system 
to be enriched with nucleosynthesis products of low-to-intermediate 
mass stars. It is clear that the currently known ultra-faint dwarf 
galaxies cannot be a dominant progenitor in metallicities 
[Fe/H]$>-2$, since they are 
more metal-poor and tend to show 
different chemical abundance patterns. Indeed, the known ultra-faint
dwarf galaxies were 
reported to have much simpler stellar population which have 
ended their major star formation events at much earlier times than 
brighter dwarf galaxies \citep{okamoto12}.
Although the surviving dSphs like Fornax or Sagittarius may 
not be a direct progenitor of these stars, the similarity in 
light elements may suggest that gas-rich systems 
that have longer star formation timescales 
have contributed to the relatively metal-rich part of 
the outer stellar halo.

\subsubsection{Comparisons with  other observations  and numerical simulations}
\label{sec:otherobs}

In order to accommodate the observed phase space and chemical abundance 
distributions, the formation of the stellar halo is 
believed to have involved multiple processes. These processes are 
frequently classified into two mechanisms, namely, a dissipational 
collapse of a proto-Galactic gas cloud 
within a very short time scale, which was introduced by a landmark 
study of \citet{eggen62}, and dissipationless 
mergers of smaller 
stellar systems with much longer timescales, which 
was proposed by \citet{searle78}. 
The existing observations of halo stars seem to favor the scenario 
that the inner part of the stellar halo is formed via 
dissipative collapse of gaseous proto-galactic fragments 
while the outer part has built up through dissipationless merging 
of small stellar systems \citep[e.g.][]{chiba00, carollo07,carollo10}.
Recently, \citet{an13} studied metallicity of the stellar halo at 
heliocentric distances in the range 5-8 kpc based on 
SDSS {\it ugriz} photometry. 
They found that the metallicity distribution function can be 
modeled with two Gaussian components with peaks at [Fe/H]$\sim -1.7$ 
and $\sim -2.3$, which favors the above scenario that at least two 
mechanisms are responsible for building up the present day stellar halo. 
It was also shown that such a hybrid scenario can naturally occur in the 
context of galaxy formation theory under the current standard cosmology
\citep[e.g.][]{bekki01,zolotov10,font11,mccarthy12,tissera12,tissera13}.

Our results show that the inner halo at relatively low metallicity 
([Fe/H]$<-2$) has similar abundance ratios as the outer halo and 
is broadly consistent with previous abundance results for the MW dSphs. 
This may suggest that, in [Fe/H]$<-2$,
inner halo stars were, at least in part, assembled from 
progenitor systems similar to these dSphs.
The presense of stars accreted from dSphs is also predected from
cosmological hydrodynamical simulation of \citet{mccarthy12}, 
which suggests that, although the inner halo is expected to be dominated 
by stars formed in situ within the proto disk and later 
dynamically heated up to the halo region,  
a small fraction of accreted populations is indeed expected. 
In the higher metallicity range, 
the abundance ratios in the inner halo subsapmle 
for some elements were clearly different from those in the thick disk 
subsample. In particular generally larger dispersion in abundance 
ratios in the inner halo stars compared to those in the thick disk stars 
 argues against the hypothesis that 
the inner halo component is entirely built up with a 
single dissipational collapse and 
star formation within a well-mixed gas. 
Rather, the results favor the senario that multiple gas-rich 
systems that have experienced 
various levels of chemical evolution prior to the merging have contributed 
to the present day inner halo.

The cosmological simulation of \citet{zolotov10} suggests that 
a certain fraction of inner halo stars are expected to form 
within the central region of a galaxy later heated up to 
halo-like orbits via mergers 
and such stars can be distinguished from accreted stars 
with their chemical abundances. 
Stars that are likely formed in such a process can indeed 
frequently be found in our inner halo subsample with 
[Fe/H]$>-2$, characterized as high [Mg/Fe], [Si/Fe], 
[Na/Fe] and [Zn/Fe] ratios.  
The simulation of \citet{mccarthy12}, suggests that the large fraction of 
the inner halo is populated by the stars formed in the 
disk system later puffed up by the dynamical heating via 
merging dark matter subhalos. 
The candidate of such population can be found in 
our sample stars having kinematics intermediate between 
the thick disk and the inner halo subsamples. These stars are 
found to have similar abundance as the thick disk 
stars inferring that a certain fraction of the halo 
is populated by stars originally belonging to the 
MW disk and later heated up while partly conserving 
their initial kinematics and chemical abundances.

Our outer halo subsample shows broadly similar
abundance ratios of light elements ($Z<30$) with many of the 
MW dSphs in [Fe/H]$<-2.0$, which
may support the hypothesis that the outer stellar halo, where 
\citet{carollo07,carollo10} found it to be dominant in [Fe/H]$<-2$, 
was assembled from stellar systems similar to the present-day 
low luminosity MW dSphs. On the other hand, our outer halo subsample 
with higher metallicity ([Fe/H]$>-2$) shows different abundance ratios 
compared to the inner halo stars. Our sample of more metal-rich outer 
halo stars is generally similar to the low-$\alpha$ stars 
reported by \citet{nissen10,nissen11} in terms of abundances 
(e.g. low [Na/Fe] and [Zn/Fe]) and kinematics (e.g. extreme retrograde orbit), 
although distinction from higher-$\alpha$ halo stars at a given [Fe/H] 
is not as clear as that observed in their works. 
These results may suggest that systems that have experienced 
extended period of chemical evolution including 
Type Ia SNe before their accretion 
may have contributed to relatively metal-rich outer halo stars 
in the solar neighborhood that have distinct 
abundances compared to the bulk of inner halo stars. 
The contribution of a certain fraction 
of the accreted stars is also suggested from medium resolution 
spectroscopic samples that  
include more distant halo stars \citep{schlaufman12,sheffield12}.

The present study of chemical abundances in the 
kinematically selected sample stars provides important insights into
the progenitor systems that gave rise to the present stellar halo 
at least in the solar neighborhood. 
However, the number of sample stars is too small and 
incomplete to quantify the fractional contribution of 
accretions to build up the stellar halo 
in the solar neighborhood.  Upcoming surveys such as {\it Gaia} and 
its spectroscopic followup will provide valuable 
insights into the origins of halo
stars by measuring accurate phase-space information and chemical 
abundances of a large number of stars within several kpc from the Sun.
Besides, kinematics and chemical abundance analysis of 
 in situ outer halo stars at Galactocentric distance 
of several tens of kpc are important 
to constrain the nature of the outer stellar halo.

\acknowledgments 

We are grateful to the referee for a careful reading of 
this manuscript and for a number of constructive 
comments, which have significantly improved our paper. 
We thank A. Tajitsu, T-S. Pyo  and the staff members of 
Subaru telescope for their helpful support and assistance in our 
HDS observation. M.N.I. is grateful to S. Wanajo, and T. Tsujimoto  
for useful discussions and comments. 
This work is supported in part from Grant-in-Aid for Scientific 
Research (23740162,23224004) of the Ministry of Education, 
Culture, Sports, Science 
and Technology in Japan.


%% in parentheses, after the keyword, but they will not be verified.

{\it Facilities:} \facility{Subaru (HDS)}.



\appendix




\clearpage

%% the file, uncomment the \plotone, \plottwo, and
%% If you need a layout that cannot be achieved with \plotone or
%% \plottwo, you can invoke the graphicx package directly with the
%% \includegraphics command or use \plotfiddle. For more information,
%% please see the tutorial on "Using Electronic Art with AASTeX" in the
%% documentation section at the AASTeX Web site,
%% http://www.journals.uchicago.edu/AAS/AASTeX.
%%
%% The examples below also include sample markup for submission of
%% supplemental electronic materials. As always, be sure to check
%% the instructions to authors for the journal you are submitting to
%% for specific submissions guidelines as they vary from

%% This example uses \plotone to include an EPS file scaled to
%% 80% of its natural size with \epsscale. Its caption
%% has been written to indicate that additional figure parts will be
%% available in the electronic journal.

\clearpage

\begin{deluxetable}{lccccccccc}
\tabletypesize{\scriptsize}
\tablecaption{ Equivalent widths \label{tab:ews}}
\tablewidth{0pt}
\tablehead{
\colhead{Object name} & \colhead{Z/Ion} & \colhead{Element} & \colhead{$\lambda$} & \colhead{$\log gf$} &
\colhead{$\chi$} & \colhead{EW} & \colhead{Flag\tablenotemark{a}} &\colhead{Refs.\tablenotemark{b}}&\colhead{HFS} \\
\colhead{} & \colhead{} & \colhead{} & \colhead{({\AA})} & \colhead{(dex)} &
\colhead{(eV)} & \colhead{({\AA})} & \colhead{} & \colhead{} &\colhead{}
}
\startdata
BD+01\arcdeg3070 & 11 1 &   NaI &    5682.63 &  $-$0.70 &   2.10 &    9.82 &  1 & NS10  &\\
BD+01\arcdeg3070 & 21 2 &  ScII &    4400.40 &  $-$0.54 &   0.61 &   67.22 &  1 & I06   & hfs\\
BD+01\arcdeg3070 & 21 2 &  ScII &    4670.42 &  $-$0.72 &   1.36 &   35.52 &  1 & R10   & hfs\\
BD+01\arcdeg3070 & 21 2 &  ScII &    5031.02 &  $-$0.40 &   1.36 &   45.74 &  1 & I06   & hfs\\
BD+01\arcdeg3070 & 21 2 &  ScII &    5239.82 &  $-$0.77 &   1.45 &   26.14 &  1 & I06   & hfs\\
\enddata
%% Text for table notes should follow after the \enddata but before
%% the \end{deluxetable}. Make sure there is at least one \tablenotemark
%% in the table for each \tablenotetext.
\tablecomments{Table \ref{tab:ews} is published in its entirety in the electronic 
edition of the Astrophysical Journal. A portion is shown here for guidance regarding its form and context.}
\tablenotetext{a}{1: Used in the abundance analysis, 0: Not used in the abundance analysis.}
\tablenotetext{b}{Reference of adopted $\log gf$. A complete list of references 
are given in the electronic version of this table.}
\end{deluxetable}

\clearpage

\begin{deluxetable}{lcccccccccc}
\rotate
\tabletypesize{\scriptsize}
\tablecaption{ Atmospheric parameters and abundances \label{tab:stpm_ab}}
\tablewidth{0pt}
\tablehead{
\colhead{Object name} & \colhead{$T_{\rm eff}$} & \colhead{$\log g$} & \colhead{$\xi$} & \colhead{[\ion{Fe}{1}/H]} & \colhead{[\ion{Fe}{2}/H]} & \colhead{[Na/Fe]} &\colhead{[Sc/Fe]} & \colhead{[V/Fe]} & \colhead{[\ion{Cr}{1}/Fe]} & \colhead{[\ion{Cr}{2}/Fe]}\\
\colhead{} & \colhead{(K)} & \colhead{(dex)} & \colhead{(km s$^{-1}$)} & \colhead{(dex)} & \colhead{(dex)} & \colhead{(dex)} &\colhead{(dex)} & \colhead{(dex)} & \colhead{(dex)} & \colhead{(dex)}
}
\startdata
BD+01\arcdeg3070 & 5404 &    3.65 &    1.18 &   $-$1.36 $\pm$   0.14 &   $-$1.37 $\pm$   0.13 &   $-$0.30 $\pm$   0.12 &    0.18 $\pm$   0.07 &    0.34 $\pm$   0.10 &   $-$0.12 $\pm$   0.06 &    0.14 $\pm$   0.08 \\
BD+04\arcdeg2466 & 5223 &    2.02 &    1.72 &   $-$1.93 $\pm$   0.14 &   $-$1.94 $\pm$   0.12 &   $-$0.08 $\pm$   0.14 &    0.13 $\pm$   0.08 &   $-$9.99 $\pm$   0.00 &   $-$0.20 $\pm$   0.08 &    0.28 $\pm$   0.10 \\
BD+04\arcdeg2621 & 4754 &    1.63 &    1.72 &   $-$2.40 $\pm$   0.16 &   $-$2.41 $\pm$   0.12 &   $-$9.99 $\pm$   0.00 &    0.08 $\pm$   0.08 &   $-$0.03 $\pm$   0.12 &   $-$0.24 $\pm$   0.09 &    0.17 $\pm$   0.07 \\
BD+09\arcdeg2870 & 4632 &    1.30 &    1.63 &   $-$2.38 $\pm$   0.17 &   $-$2.42 $\pm$   0.12 &   $-$9.99 $\pm$   0.00 &    0.07 $\pm$   0.08 &   $-$0.14 $\pm$   0.11 &   $-$0.31 $\pm$   0.08 &    0.11 $\pm$   0.07 \\
BD+10\arcdeg2495 & 4974 &    2.29 &    1.64 &   $-$2.01 $\pm$   0.15 &   $-$2.01 $\pm$   0.12 &   $-$9.99 $\pm$   0.00 &    0.02 $\pm$   0.07 &   $-$0.04 $\pm$   0.10 &   $-$0.21 $\pm$   0.07 &    0.17 $\pm$   0.07 \\
\enddata
%% Text for table notes should follow after the \enddata but before
%% the \end{deluxetable}. Make sure there is at least one \tablenotemark
%% in the table for each \tablenotetext.
\tablecomments{Table \ref{tab:stpm_ab} is published in its 
entirety in the electronic edition of the Astrophysical Journal. 
A portion is shown here for guidance regarding its form and context.}
%\tablenotetext{a}{  }
\end{deluxetable}

\clearpage

\begin{deluxetable}{lcccccccccccccc}
\rotate
\tabletypesize{\scriptsize}
\tablecaption{Comparison with \citet{nissen10,nissen11}\label{tab:comp_nissen}}
\tablewidth{0pt}
\tablehead{
\colhead{Starname}&\colhead{NS10/TW} & \colhead{$T_{\rm eff}$} & \colhead{$\log g$} & \colhead{$\xi$} & \colhead{[Fe/H] } &\colhead{[Na/Fe]} & \colhead{[\ion{Cr}{1}/Fe]} & \colhead{[Mn/Fe]} &\colhead{[Ni/Fe]} & \colhead{[Cu/Fe]}& \colhead{[Zn/Fe]} & \colhead{[Y/Fe]} & \colhead{[Ba/Fe]} & \colhead{Classification}\\
\colhead{} & \colhead{}&\colhead{(K)} & \colhead{(dex)} & \colhead{(km s$^{-1}$)} & \colhead{(dex)} &\colhead{(dex)} & \colhead{(dex)} & \colhead{(dex)} & \colhead{(dex)} &\colhead{(dex)} & \colhead{(dex)}& \colhead{(dex)} & \colhead{(dex)} & \colhead{}
}
\startdata
         G112-43 & TW &$ 6176$ & $  4.0$ & $  1.4$ & $  -1.33$ & $  -0.11$ & $  -0.04$ & $  -0.20$ & $   0.00$ & $  -0.49$ & $   0.30$ & $  -0.16$ & $  -0.21$ & OH \\
      & NS &$ 6074$ & $  4.0$ & $  1.3$ & $  -1.25$ & $  -0.11$ & $   0.00$ & $  -0.19$ & $  -0.02$ & $  -0.52$ & $   0.30$ & $  -0.14$ & $  -0.27$ & low-alpha \\
          G53-41 & TW &$ 6070$ & $  4.6$ & $  0.8$ & $  -1.15$ & $   0.10$ & $  -0.09$ & $  -0.32$ & $  -0.14$ & $  -0.75$ & $   0.04$ & $   0.10$ & $   0.54$ & IH \\
      & NS &$ 5859$ & $  4.3$ & $  1.3$ & $  -1.20$ & $   0.23$ & $  -0.03$ & $  -0.38$ & $  -0.09$ &   \nodata & $   0.03$ & $   0.05$ & $   0.24$ & low-alpha \\
         G125-13 & TW &$ 6079$ & $  4.8$ & $  0.8$ & $  -1.35$ & $  -0.20$ & $  -0.14$ & $  -0.32$ & $  -0.07$ & $  -0.80$ & $   0.03$ & $  -0.10$ & $  -0.01$ & IH \\
      & NS &$ 5848$ & $  4.3$ & $  1.5$ & $  -1.43$ & $  -0.17$ & $  -0.09$ & $  -0.39$ & $  -0.09$ &   \nodata & $   0.05$ & $  -0.22$ & $  -0.31$ & (high-alpha) \\
        HD111980 & TW &$ 5798$ & $  4.0$ & $  1.2$ & $  -1.13$ & $   0.12$ & $  -0.14$ & $  -0.30$ & $   0.00$ &   \nodata & $   0.15$ & $   0.23$ & $   0.32$ & OH \\
      & NS &$ 5778$ & $  4.0$ & $  1.5$ & $  -1.08$ & $   0.03$ & $  -0.02$ & $  -0.34$ & $   0.00$ & $  -0.32$ & $   0.15$ & $   0.16$ & $   0.07$ & high-alpha \\
          G20-15 & TW &$ 6042$ & $  4.3$ & $  1.2$ & $  -1.62$ &   \nodata & $  -0.17$ & $  -0.26$ & $  -0.01$ &   \nodata & $   0.11$ & $  -0.16$ & $   0.02$ & OH \\
      & NS &$ 6027$ & $  4.3$ & $  1.6$ & $  -1.49$ & $  -0.18$ & $  -0.03$ & $  -0.32$ & $  -0.05$ &   \nodata & $   0.11$ & $  -0.14$ & $  -0.22$ & (low-alpha) \\
        HD105004 & TW &$ 6115$ & $  5.0$ & $  0.4$ & $  -0.60$ & $  -0.09$ & $  -0.08$ & $  -0.12$ & $  -0.12$ &   \nodata & $   0.01$ & $   0.02$ & $   0.21$ & IH \\
      & NS &$ 5754$ & $  4.3$ & $  1.2$ & $  -0.82$ & $  -0.05$ & $  -0.03$ & $  -0.21$ & $  -0.06$ & $  -0.22$ & $   0.04$ & $  -0.15$ & $  -0.16$ & low-alpha \\
         G176-53 & TW &$ 5753$ & $  5.0$ & $  0.2$ & $  -1.26$ & $  -0.30$ & $  -0.06$ & $  -0.31$ & $  -0.08$ &   \nodata & $   0.04$ & $  -0.05$ & $   0.12$ & OH \\
      & NS &$ 5523$ & $  4.5$ & $  1.0$ & $  -1.34$ & $  -0.36$ & $  -0.01$ & $  -0.35$ & $  -0.12$ & $  -0.57$ & $   0.08$ & $  -0.26$ & $  -0.26$ & low-alpha \\
        HD193901 & TW &$ 5908$ & $  4.9$ & $  0.3$ & $  -0.95$ & $  -0.28$ & $  -0.05$ & $  -0.18$ & $  -0.13$ & $  -0.55$ & $  -0.04$ & $  -0.04$ & $   0.15$ & IH \\
      & NS &$ 5650$ & $  4.4$ & $  1.2$ & $  -1.09$ & $  -0.27$ & $  -0.04$ & $  -0.34$ & $  -0.14$ & $  -0.62$ & $  -0.03$ & $  -0.20$ & $  -0.19$ & low-alpha \\
         G188-22 & TW &$ 6170$ & $  4.5$ & $  1.1$ & $  -1.29$ & $   0.02$ & $  -0.09$ & $  -0.27$ & $  -0.02$ &   \nodata & $   0.11$ & $   0.26$ & $   0.32$ & IH/TD \\
      & NS &$ 5974$ & $  4.2$ & $  1.5$ & $  -1.32$ & $  -0.04$ & $  -0.04$ & $  -0.34$ & $  -0.01$ & $  -0.41$ & $   0.15$ & $   0.15$ & $   0.03$ & high-alpha \\
\enddata
%% Text for table notes should follow after the \enddata but before
%% the \end{deluxetable}. Make sure there is at least one \tablenotemark
%% in the table for each \tablenotetext.
%\tablecomments{}
%\tablenotetext{a}{  }
\end{deluxetable}

\clearpage

\begin{deluxetable}{lcccccccccccccc}
\rotate
\tabletypesize{\scriptsize}
\tablecaption{Comparison with \citet{roederer10}\label{tab:comp_roederer}}
\tablewidth{0pt}
\tablehead{
\colhead{Name}&\colhead{$T_{\rm eff,TW}$} & \colhead{$T_{\rm eff,R10}$} & \colhead{$\log g_{\rm TW}$} & \colhead{$\log g_{\rm R10}$} & \colhead{$\xi_{\rm TW}$} &\colhead{$\xi_{\rm R10}$} & \colhead{[Fe/H]$_{\rm TW}$} & \colhead{[Fe/H]$_{\rm TW}$} & \colhead{[Zn/Fe]$_{\rm TW}$}& \colhead{[Zn/Fe]$_{\rm R10}$} & \colhead{[Y/Fe]$_{\rm TW}$} & \colhead{[Y/Fe]$_{\rm R10}$}& \colhead{[Eu/Fe]$_{\rm TW}$}& \colhead{[Eu/Fe]$_{\rm R10}$} \\
\colhead{} & \multicolumn{2}{c}{(K)} & \multicolumn{2}{c}{(dex)}& \multicolumn{2}{c}{(km s$^{-1}$)} & \multicolumn{2}{c}{(dex)} &\multicolumn{2}{c}{(dex)} & \multicolumn{2}{c}{(dex)}& \multicolumn{2}{c}{(dex)}
}
\startdata
        HD107752 & $ 4826$ & $ 4649$ & $  1.6$ & $  1.6$ & $  1.9$ & $  2.0$ & $  -2.78$ & $  -2.78$ & $   1.92$ & $   1.93$ & $  -0.84$ & $  -0.90$ & $  -1.90$ & $  -1.99$ \\
        HD119516 & $ 5605$ & $ 5382$ & $  1.9$ & $  2.5$ & $  1.9$ & $  2.5$ & $  -1.92$ & $  -2.26$ & $   2.67$ & $   2.32$ & $   0.02$ & $  -0.43$ & $  -0.98$ & $  -1.43$ \\
        HD124358 & $ 4745$ & $ 4688$ & $  1.4$ & $  1.6$ & $  1.8$ & $  2.1$ & $  -1.70$ & $  -1.91$ & $   2.73$ & $   2.64$ & $  -0.13$ & $  -0.22$ & $  -1.02$ & $  -0.94$ \\
        HD128279 & $ 5328$ & $ 5290$ & $  3.2$ & $  3.0$ & $  1.4$ & $  1.5$ & $  -2.18$ & $  -2.51$ & $   2.45$ & $   2.15$ &   \nodata & $  -1.04$ & $  -1.60$ & $  -2.27$ \\
         HD85773 & $ 4370$ & $ 4268$ & $  0.7$ & $  0.5$ & $  2.0$ & $  2.0$ & $  -2.44$ & $  -2.62$ & $   2.65$ & $   2.56$ & $  -0.80$ & $  -0.93$ & $  -1.95$ & $  -1.84$ \\
          G17-25 & $ 5174$ & $ 4966$ & $  4.8$ & $  4.3$ & $  0.0$ & $  0.8$ & $  -1.14$ & $  -1.54$ & $   3.68$ & $   3.40$ &   \nodata & $   0.87$ & $  -0.33$ & $   0.00$ \\
        HD214362 & $ 5783$ & $ 5727$ & $  2.0$ & $  2.6$ & $  2.3$ & $  2.0$ & $  -1.90$ & $  -1.87$ & $   2.76$ & $   2.71$ & $   0.22$ & $   0.32$ & $  -1.17$ & $  -0.82$ \\
        HD218857 & $ 5107$ & $ 5103$ & $  2.7$ & $  2.4$ & $  1.7$ & $  1.9$ & $  -1.91$ & $  -1.90$ & $   2.75$ & $   2.64$ & $  -0.09$ & $  -0.19$ & $  -1.55$ & $  -1.42$ \\
         G153-21 & $ 5566$ & $ 5700$ & $  3.9$ & $  4.4$ & $  0.9$ & $  1.4$ & $  -0.65$ & $  -0.70$ & $   4.16$ & $   4.06$ & $   1.36$ & $   1.46$ & $   0.01$ & $   0.34$ \\
         G176-53 & $ 5753$ & $ 5593$ & $  5.0$ & $  4.5$ & $  0.2$ & $  1.2$ & $  -1.26$ & $  -1.34$ & $   3.38$ & $   3.18$ & $   0.90$ & $   0.63$ & $  -0.21$ & $  -0.32$ \\
         G188-22 & $ 6170$ & $ 5827$ & $  4.5$ & $  4.3$ & $  1.1$ & $  1.2$ & $  -1.29$ & $  -1.52$ & $   3.41$ & $   3.24$ & $   1.18$ & $   0.94$ & $  -0.44$ & $  -0.60$ \\
          G63-46 & $ 5867$ & $ 5705$ & $  4.6$ & $  4.2$ & $  0.6$ & $  1.3$ & $  -0.62$ & $  -0.90$ & $   4.12$ & $   3.86$ & $   1.70$ & $   1.26$ & $   0.20$ & $  -0.05$ \\
          G23-14 & $ 5061$ & $ 5025$ & $  3.1$ & $  3.0$ & $  1.1$ & $  1.3$ & $  -1.47$ & $  -1.64$ & $   3.19$ & $   3.05$ & $   0.61$ & $   0.46$ & $  -0.52$ & $  -0.58$ \\
        HD105546 & $ 5179$ & $ 5190$ & $  2.3$ & $  2.5$ & $  1.8$ & $  1.6$ & $  -1.44$ & $  -1.48$ & $   3.24$ & $   3.29$ & $   0.76$ & $   0.74$ & $  -0.63$ & $  -0.56$ \\
        HD108317 & $ 5284$ & $ 5234$ & $  2.9$ & $  2.7$ & $  1.5$ & $  2.0$ & $  -2.27$ & $  -2.18$ & $   2.43$ & $   2.40$ & $  -0.22$ & $  -0.39$ & $  -1.22$ & $  -1.32$ \\
        HD122956 & $ 4609$ & $ 4508$ & $  1.6$ & $  1.6$ & $  1.7$ & $  1.6$ & $  -1.71$ & $  -1.95$ & $   2.92$ & $   2.87$ & $   0.29$ & $   0.16$ & $  -0.83$ & $  -0.79$ \\
        HD171496 & $ 4795$ & $ 4952$ & $  1.9$ & $  2.4$ & $  1.1$ & $  1.4$ & $  -0.64$ & $  -0.67$ & $   4.23$ & $   4.11$ & $   1.22$ & $   1.40$ & $  -0.23$ & $   0.11$ \\
        HD184266 & $ 5618$ & $ 6000$ & $  1.6$ & $  2.7$ & $  2.4$ & $  3.0$ & $  -1.68$ & $  -1.43$ & $   2.99$ & $   3.19$ & $   0.22$ & $   0.69$ & $  -1.08$ & $  -0.43$ \\
        HD188510 & $ 5654$ & $ 5564$ & $  5.0$ & $  4.5$ & $  0.1$ & $  1.0$ & $  -1.47$ & $  -1.32$ & $   3.15$ & $   3.01$ & $   0.71$ & $   0.44$ & $  -0.55$ & $  -0.52$ \\
        HD193901 & $ 5908$ & $ 5750$ & $  4.9$ & $  4.5$ & $  0.3$ & $  1.5$ & $  -0.95$ & $  -1.08$ & $   3.59$ & $   3.36$ & $   1.21$ & $   0.83$ & $   0.06$ & $  -0.10$ \\
        HD210295 & $ 4763$ & $ 4750$ & $  2.2$ & $  2.5$ & $  1.3$ & $  1.6$ & $  -1.24$ & $  -1.46$ & $   3.52$ & $   3.37$ & $   1.00$ & $   0.85$ & $  -0.46$ & $  -0.34$ \\
\enddata
%% Text for table notes should follow after the \enddata but before
%% the \end{deluxetable}. Make sure there is at least one \tablenotemark
%% in the table for each \tablenotetext.
%\tablecomments{}
%\tablenotetext{a}{  }
\end{deluxetable}

\clearpage

{\tabcolsep=1.7mm
\begin{deluxetable}{lcccccccccccccccc}
\rotate
\tabletypesize{\tiny}
\tablecaption{ Means and standard deviations in the abundance ratios \label{tab:mean_dev}}
\tablewidth{0pt}
\tablehead{
\colhead{[X/Fe]} & \colhead{}&\multicolumn{5}{c}{[Fe/H]$>-1.5$} & \multicolumn{5}{c}{$-2.5<$[Fe/H]$\leq-1.5$} & \multicolumn{5}{c}{[Fe/H]$\leq -2.5$} \\ \cline{3-17}
\colhead{} &\colhead{} & \colhead{$\mu$\tablenotemark{a}}& \colhead{$\sigma$\tablenotemark{b}} & \colhead{N\tablenotemark{c}} & \colhead{$\mu_{\rm d}$(N)\tablenotemark{d}} & \colhead{$\mu_{\rm g}$(N)\tablenotemark{e}}& \colhead{$\mu$}& \colhead{$\sigma$} & \colhead{N} &\colhead{$\mu_{\rm d}$(N)} &  \colhead{$\mu_{\rm g}$(N)}& \colhead{$\mu$} & \colhead{$\sigma$} & \colhead{N} & \colhead{$\mu_{\rm d}$(N)}& \colhead{$\mu_{\rm g}$(N)}  
}
\startdata

   Na      & TD &    0.10 &    0.11 &  8 &    0.16$\pm$0.06 (2) &    0.13$\pm$0.04 (4) &   $-$0.08 &    0.08 &  2 &       \nodata        &   $-$0.03$\pm$0.14 (1) &  \multicolumn{3}{c}{\nodata} &       \nodata        &       \nodata        \\
           & IH &   $-$0.13 &    0.17 & 16 &   $-$0.14$\pm$0.02 (11) &   $-$0.13$\pm$0.07 (3) &   $-$0.15 &    0.14 &  8 &   $-$0.19$\pm$0.05 (4) &   $-$0.11$\pm$0.04 (2) &  \multicolumn{3}{c}{\nodata} &       \nodata        &       \nodata        \\
           & OH &   $-$0.28 &    0.19 & 11 &   $-$0.23$\pm$0.03 (6) &   $-$0.30$\pm$0.14 (1) &   $-$0.28 &    0.18 &  5 &   $-$0.01$\pm$0.09 (1) &   $-$0.35$\pm$0.02 (4) &   $-$0.06 &    0.14 &  1 &       \nodata        &   $-$0.06$\pm$0.14 (1) \\
   Sc      & TD &    0.21 &    0.09 &  8 &    0.29$\pm$0.07 (2) &    0.22$\pm$0.02 (4) &    0.15 &    0.05 &  2 &       \nodata        &    0.19$\pm$0.08 (1) &    0.17 &    0.10 &  1 &       \nodata        &       \nodata        \\
           & IH &    0.14 &    0.07 & 18 &    0.15$\pm$0.02 (12) &    0.15$\pm$0.06 (3) &    0.13 &    0.09 & 13 &    0.18$\pm$0.04 (6) &    0.07$\pm$0.02 (4) &    0.14 &    0.26 &  3 &    0.28$\pm$0.04 (2) &   $-$0.16$\pm$0.08 (1) \\
           & OH &    0.14 &    0.06 & 11 &    0.16$\pm$0.01 (6) &   $-$0.03$\pm$0.12 (1) &    0.05 &    0.10 & 20 &    0.08$\pm$0.05 (5) &    0.03$\pm$0.02 (8) &    0.24 &    0.26 &  6 &    0.36$\pm$0.11 (4) &    0.04$\pm$0.09 (1) \\
    V      & TD &    0.02 &    0.19 &  8 &    0.22$\pm$0.11 (2) &   $-$0.10$\pm$0.08 (4) &    0.04 &    0.02 &  2 &       \nodata        &    0.05$\pm$0.13 (1) &  \multicolumn{3}{c}{\nodata} &       \nodata        &       \nodata        \\
           & IH &    0.06 &    0.13 & 15 &    0.12$\pm$0.04 (9) &   $-$0.11$\pm$0.03 (3) &    0.14 &    0.24 &  8 &    0.28$\pm$0.08 (5) &   $-$0.09$\pm$0.05 (3) &    0.17 &    0.17 &  1 &    0.17$\pm$0.17 (1) &       \nodata        \\
           & OH &    0.12 &    0.12 & 10 &    0.15$\pm$0.02 (6) &   $-$0.12$\pm$0.17 (1) &   $-$0.05 &    0.13 & 11 &   $-$0.19$\pm$0.13 (1) &   $-$0.09$\pm$0.02 (6) &   $-$0.27 &    0.16 &  1 &       \nodata        &   $-$0.27$\pm$0.16 (1) \\
 Cr I      & TD &   $-$0.13 &    0.11 &  8 &   $-$0.07$\pm$0.00 (2) &   $-$0.14$\pm$0.07 (4) &   $-$0.23 &    0.07 &  2 &       \nodata        &   $-$0.18$\pm$0.07 (1) &   $-$0.06 &    0.07 &  1 &       \nodata        &       \nodata        \\
           & IH &   $-$0.14 &    0.12 & 18 &   $-$0.09$\pm$0.01 (12) &   $-$0.29$\pm$0.11 (3) &   $-$0.14 &    0.08 & 13 &   $-$0.07$\pm$0.01 (6) &   $-$0.22$\pm$0.02 (4) &   $-$0.06 &    0.10 &  3 &   $-$0.00$\pm$0.01 (2) &   $-$0.17$\pm$0.14 (1) \\
           & OH &   $-$0.10 &    0.07 & 11 &   $-$0.08$\pm$0.01 (6) &   $-$0.26$\pm$0.18 (1) &   $-$0.17 &    0.09 & 20 &   $-$0.12$\pm$0.03 (5) &   $-$0.26$\pm$0.02 (8) &   $-$0.23 &    0.12 &  5 &   $-$0.18$\pm$0.08 (3) &   $-$0.31$\pm$0.08 (1) \\
 Cr II     & TD &    0.18 &    0.06 &  8 &    0.22$\pm$0.01 (2) &    0.15$\pm$0.03 (4) &    0.18 &    0.06 &  2 &       \nodata        &    0.22$\pm$0.09 (1) &    0.18 &    0.10 &  1 &       \nodata        &       \nodata        \\
           & IH &    0.16 &    0.05 & 18 &    0.17$\pm$0.01 (12) &    0.13$\pm$0.05 (3) &    0.19 &    0.05 & 13 &    0.20$\pm$0.02 (6) &    0.17$\pm$0.02 (4) &    0.16 &    0.07 &  1 &       \nodata        &    0.16$\pm$0.07 (1) \\
           & OH &    0.17 &    0.05 & 11 &    0.17$\pm$0.02 (6) &    0.09$\pm$0.09 (1) &    0.15 &    0.06 & 19 &    0.15$\pm$0.02 (5) &    0.13$\pm$0.02 (8) &    0.08 &    0.06 &  1 &       \nodata        &    0.08$\pm$0.06 (1) \\
  Mn       & TD &   $-$0.08 &    0.18 &  8 &   $-$0.03$\pm$0.10 (2) &    0.01$\pm$0.07 (4) &   $-$0.33 &    0.08 &  2 &       \nodata        &   $-$0.27$\pm$0.07 (1) &   $-$0.68 &    0.07 &  1 &       \nodata        &       \nodata        \\
           & IH &   $-$0.25 &    0.08 & 18 &   $-$0.27$\pm$0.02 (12) &   $-$0.22$\pm$0.01 (3) &   $-$0.34 &    0.10 & 13 &   $-$0.35$\pm$0.02 (6) &   $-$0.23$\pm$0.04 (4) &   $-$0.44 &    0.12 &  2 &   $-$0.35$\pm$0.16 (1) &   $-$0.53$\pm$0.14 (1) \\
           & OH &   $-$0.27 &    0.08 & 11 &   $-$0.28$\pm$0.02 (6) &   $-$0.17$\pm$0.08 (1) &   $-$0.37 &    0.10 & 20 &   $-$0.34$\pm$0.03 (5) &   $-$0.42$\pm$0.03 (8) &   $-$0.34 &    0.16 &  1 &       \nodata        &   $-$0.34$\pm$0.16 (1) \\
  Co       & TD &   $-$0.05 &    0.12 &  5 &   $-$0.07$\pm$0.13 (1) &   $-$0.04$\pm$0.07 (4) &    0.12 &    0.19 &  2 &       \nodata        &   $-$0.02$\pm$0.14 (1) &    0.05 &    0.10 &  1 &       \nodata        &       \nodata        \\
           & IH &    0.08 &    0.15 &  5 &    0.20$\pm$0.03 (2) &    0.04$\pm$0.12 (2) &    0.11 &    0.13 &  8 &    0.13$\pm$0.03 (5) &    0.09$\pm$0.13 (3) &    0.17 &    0.03 &  2 &    0.15$\pm$0.16 (1) &    0.19$\pm$0.17 (1) \\
           & OH &    0.10 &    0.10 &  5 &    0.14$\pm$0.02 (4) &   $-$0.05$\pm$0.16 (1) &    0.06 &    0.13 &  9 &    0.10$\pm$0.01 (4) &    0.32$\pm$0.22 (1) &  \multicolumn{3}{c}{\nodata} &       \nodata        &       \nodata        \\
  Ni       & TD &   $-$0.01 &    0.05 &  8 &    0.02$\pm$0.02 (2) &    0.00$\pm$0.01 (4) &   $-$0.06 &    0.05 &  2 &       \nodata        &   $-$0.10$\pm$0.10 (1) &  \multicolumn{3}{c}{\nodata} &       \nodata        &       \nodata        \\
           & IH &   $-$0.08 &    0.05 & 18 &   $-$0.09$\pm$0.01 (12) &   $-$0.08$\pm$0.01 (3) &   $-$0.08 &    0.06 & 12 &   $-$0.08$\pm$0.04 (5) &   $-$0.07$\pm$0.03 (4) &   $-$0.06 &    0.04 &  2 &   $-$0.09$\pm$0.18 (1) &   $-$0.04$\pm$0.11 (1) \\
           & OH &   $-$0.12 &    0.07 & 11 &   $-$0.08$\pm$0.03 (6) &   $-$0.17$\pm$0.09 (1) &   $-$0.10 &    0.09 & 19 &   $-$0.14$\pm$0.04 (5) &   $-$0.13$\pm$0.02 (8) &    0.03 &    0.12 &  2 &    0.12$\pm$0.08 (1) &   $-$0.06$\pm$0.08 (1) \\
  Cu       & TD &   $-$0.26 &    0.23 &  7 &   $-$0.07$\pm$0.01 (2) &   $-$0.25$\pm$0.05 (4) &   $-$0.57 &    0.12 &  1 &       \nodata        &   $-$0.57$\pm$0.12 (1) &  \multicolumn{3}{c}{\nodata} &       \nodata        &       \nodata        \\
           & IH &   $-$0.49 &    0.29 & 11 &   $-$0.53$\pm$0.09 (6) &   $-$0.20$\pm$0.27 (2) &   $-$0.70 &    0.24 &  4 &   $-$0.45$\pm$0.12 (1) &   $-$0.67$\pm$0.08 (2) &  \multicolumn{3}{c}{\nodata} &       \nodata        &       \nodata        \\
           & OH &   $-$0.54 &    0.18 &  6 &   $-$0.49$\pm$0.12 (1) &   $-$0.85$\pm$0.12 (1) &   $-$0.74 &    0.27 &  8 &       \nodata        &   $-$0.87$\pm$0.07 (6) &   $-$0.68 &    0.12 &  1 &       \nodata        &   $-$0.68$\pm$0.12 (1) \\
  Zn       & TD &    0.20 &    0.08 &  8 &    0.22$\pm$0.04 (2) &    0.22$\pm$0.04 (4) &    0.14 &    0.00 &  2 &       \nodata        &    0.14$\pm$0.18 (1) &  \multicolumn{3}{c}{\nodata} &       \nodata        &       \nodata        \\
           & IH &    0.06 &    0.09 & 18 &    0.03$\pm$0.02 (12) &    0.14$\pm$0.06 (3) &    0.13 &    0.15 & 12 &    0.04$\pm$0.04 (5) &    0.23$\pm$0.10 (4) &   $-$0.01 &    0.27 &  2 &   $-$0.20$\pm$0.21 (1) &    0.18$\pm$0.13 (1) \\
           & OH &    0.03 &    0.12 & 11 &    0.07$\pm$0.05 (6) &   $-$0.07$\pm$0.19 (1) &    0.01 &    0.10 & 19 &    0.01$\pm$0.06 (5) &    0.01$\pm$0.04 (8) &    0.11 &    0.10 &  2 &       \nodata        &    0.04$\pm$0.21 (1) \\
  Sr       & TD &  \multicolumn{3}{c}{\nodata} &       \nodata        &       \nodata        &   $-$0.03 &    0.23 &  1 &       \nodata        &       \nodata        &  \multicolumn{3}{c}{\nodata} &       \nodata        &       \nodata        \\
           & IH &   $-$0.07 &    0.10 &  6 &   $-$0.07$\pm$0.04 (6) &       \nodata        &    0.13 &    0.15 &  5 &    0.13$\pm$0.07 (5) &       \nodata        &    0.09 &    0.17 &  3 &    0.16$\pm$0.12 (2) &   $-$0.04$\pm$0.18 (1) \\
           & OH &    0.02 &    0.07 &  4 &    0.02$\pm$0.03 (4) &       \nodata        &   $-$0.09 &    0.25 &  6 &   $-$0.04$\pm$0.11 (5) &       \nodata        &    0.06 &    0.12 &  4 &    0.08$\pm$0.08 (3) &       \nodata        \\
   Y       & TD &   $-$0.11 &    0.14 &  8 &   $-$0.05$\pm$0.16 (2) &   $-$0.16$\pm$0.07 (4) &   $-$0.10 &    0.09 &  2 &       \nodata        &   $-$0.04$\pm$0.09 (1) &   $-$0.04 &    0.09 &  1 &       \nodata        &       \nodata        \\
           & IH &    0.06 &    0.20 & 17 &    0.09$\pm$0.06 (12) &    0.07$\pm$0.03 (3) &   $-$0.12 &    0.32 & 11 &    0.02$\pm$0.10 (4) &   $-$0.36$\pm$0.09 (4) &   $-$0.07 &    0.27 &  2 &    0.12$\pm$0.12 (1) &   $-$0.26$\pm$0.09 (1) \\
           & OH &   $-$0.02 &    0.12 &  9 &   $-$0.00$\pm$0.05 (6) &       \nodata        &   $-$0.23 &    0.20 & 16 &   $-$0.14$\pm$0.12 (4) &   $-$0.32$\pm$0.06 (8) &   $-$0.46 &    0.05 &  2 &       \nodata        &   $-$0.49$\pm$0.10 (1) \\
 Zr        & TD &    0.12 &    0.21 &  4 &    0.10$\pm$0.20 (2) &   $-$0.04$\pm$0.10 (1) &    0.25 &    0.01 &  2 &       \nodata        &    0.25$\pm$0.11 (1) &    0.42 &    0.09 &  1 &       \nodata        &       \nodata        \\
           & IH &    0.29 &    0.18 &  4 &    0.40$\pm$0.15 (2) &       \nodata        &    0.20 &    0.37 &  6 &    0.40$\pm$0.24 (3) &    0.01$\pm$0.15 (3) &    0.20 &    0.09 &  1 &       \nodata        &    0.20$\pm$0.09 (1) \\
           & OH &    0.34 &    0.12 &  5 &    0.32$\pm$0.06 (4) &       \nodata        &   $-$0.06 &    0.11 &  7 &       \nodata        &   $-$0.03$\pm$0.04 (6) &   $-$0.16 &    0.14 &  1 &       \nodata        &   $-$0.16$\pm$0.14 (1) \\
  Ba       & TD &    0.09 &    0.07 &  8 &    0.05$\pm$0.06 (2) &    0.13$\pm$0.03 (4) &    0.06 &    0.08 &  2 &       \nodata        &    0.12$\pm$0.22 (1) &   $-$0.50 &    0.12 &  1 &       \nodata        &       \nodata        \\
           & IH &    0.21 &    0.29 & 18 &    0.27$\pm$0.10 (12) &    0.07$\pm$0.07 (3) &   $-$0.03 &    0.53 & 13 &    0.00$\pm$0.08 (6) &   $-$0.34$\pm$0.20 (4) &   $-$0.25 &    0.14 &  3 &   $-$0.24$\pm$0.14 (2) &   $-$0.25$\pm$0.12 (1) \\
           & OH &    0.09 &    0.15 & 10 &    0.11$\pm$0.07 (6) &       \nodata        &   $-$0.11 &    0.26 & 20 &    0.03$\pm$0.08 (5) &   $-$0.23$\pm$0.09 (8) &   $-$0.28 &    0.37 &  5 &   $-$0.03$\pm$0.05 (3) &   $-$0.46$\pm$0.17 (1) \\
  La       & TD &    0.14 &    0.08 &  5 &       \nodata        &    0.14$\pm$0.05 (4) &    0.31 &    0.11 &  2 &       \nodata        &    0.39$\pm$0.15 (1) &  \multicolumn{3}{c}{\nodata} &       \nodata        &       \nodata        \\
           & IH &    0.42 &    0.48 &  9 &    0.73$\pm$0.29 (4) &    0.15$\pm$0.09 (3) &    0.43 &    0.63 &  5 &       \nodata        &    0.16$\pm$0.08 (2) &  \multicolumn{3}{c}{\nodata} &       \nodata        &       \nodata        \\
           & OH &    0.46 &    0.35 &  6 &    0.36$\pm$0.12 (3) &    0.25$\pm$0.14 (1) &    0.34 &    0.47 &  7 &    0.36$\pm$0.14 (1) &    0.12$\pm$0.09 (4) &   $-$0.02 &    0.14 &  1 &       \nodata        &   $-$0.02$\pm$0.14 (1) \\
 Nd        & TD &    0.27 &    0.19 &  8 &    0.55$\pm$0.15 (2) &    0.16$\pm$0.01 (4) &    0.33 &    0.00 &  2 &       \nodata        &    0.33$\pm$0.10 (1) &  \multicolumn{3}{c}{\nodata} &       \nodata        &       \nodata        \\
           & IH &    0.57 &    0.30 & 18 &    0.68$\pm$0.07 (12) &    0.23$\pm$0.01 (3) &    0.30 &    0.53 & 10 &    0.62$\pm$0.16 (3) &   $-$0.12$\pm$0.18 (4) &    0.17 &    0.10 &  1 &       \nodata        &    0.17$\pm$0.10 (1) \\
           & OH &    0.50 &    0.17 & 10 &    0.62$\pm$0.07 (5) &    0.35$\pm$0.14 (1) &    0.23 &    0.26 & 14 &    0.57$\pm$0.04 (3) &    0.09$\pm$0.05 (7) &   $-$0.11 &    0.11 &  1 &       \nodata        &   $-$0.11$\pm$0.11 (1) \\
  Sm       & TD &    0.19 &    0.11 &  8 &    0.21$\pm$0.10 (2) &    0.14$\pm$0.05 (4) &    0.50 &    0.07 &  2 &       \nodata        &    0.45$\pm$0.10 (1) &  \multicolumn{3}{c}{\nodata} &       \nodata        &       \nodata        \\
           & IH &    0.50 &    0.34 & 10 &    0.82$\pm$0.16 (4) &    0.21$\pm$0.06 (3) &    0.35 &    0.50 &  7 &       \nodata        &    0.11$\pm$0.12 (4) &    0.33 &    0.11 &  1 &       \nodata        &    0.33$\pm$0.11 (1) \\
           & OH &    0.60 &    0.16 &  7 &    0.68$\pm$0.13 (3) &    0.47$\pm$0.16 (1) &    0.26 &    0.13 & 10 &       \nodata        &    0.25$\pm$0.05 (7) &    0.03 &    0.12 &  1 &       \nodata        &    0.03$\pm$0.12 (1) \\
  Eu       & TD &    0.19 &    0.16 &  8 &    0.22$\pm$0.09 (2) &    0.09$\pm$0.07 (4) &    0.51 &    0.03 &  2 &       \nodata        &    0.48$\pm$0.10 (1) &    0.06 &    0.21 &  1 &       \nodata        &       \nodata        \\
           & IH &    0.39 &    0.27 & 18 &    0.48$\pm$0.04 (12) &    0.01$\pm$0.25 (3) &    0.20 &    0.42 &  9 &    0.35$\pm$0.12 (5) &   $-$0.01$\pm$0.23 (4) &    0.43 &    0.10 &  2 &    0.70$\pm$0.20 (2) &    0.36$\pm$0.21 (1) \\
           & OH &    0.50 &    0.21 & 10 &    0.49$\pm$0.08 (6) &    0.23$\pm$0.21 (1) &    0.34 &    0.32 & 20 &    0.59$\pm$0.06 (5) &    0.16$\pm$0.11 (8) &    0.09 &    0.11 &  1 &    1.07$\pm$0.31 (2) &    0.09$\pm$0.11 (1) \\
\enddata
%% Text for table notes should follow after the \enddata but before
%% the \end{deluxetable}. Make sure there is at least one \tablenotemark
%% in the table for each \tablenotetext.
%\tablecomments{   }
\tablenotetext{a}{Means of the abundance ratios within a given [Fe/H] interval.}
\tablenotetext{b}{Standard deviations of the means.}
\tablenotetext{c}{The number of stars used to calculate the $\mu$ and $\sigma$.}
\tablenotetext{d}{A mean and its error for dwarf stars only. The number of stars is 
given in parenthesis.}
\tablenotetext{e}{A mean and its error for giant stars only. The number of stars is 
given in parenthesis.}
\end{deluxetable}
}

%% sample file. (Note that you will need to comment out the \documentclass,
%% \begin{document}, and \end{document} commands from table.tex if you want
%% to include it in this document.)


%% The following command ends your manuscript. LaTeX will ignore any text

\end{document}